\documentclass[12pt]{article}

\usepackage{bbm}
\usepackage{amssymb,amsmath,amsthm}

\usepackage[dvips,xdvi]{graphics}
\usepackage[dvips,xdvi]{graphicx}

\newcommand{\llhd}{\,\lhd\,}
\newcommand{\saa}{\Sigma( 36 \hspace{-2.5pt} \times \hspace{-2.5pt} 3)}
\newcommand{\sbb}{\Sigma( 72 \hspace{-2.5pt} \times \hspace{-2.5pt} 3)}
\newcommand{\scc}{\Sigma(216 \hspace{-2.5pt} \times \hspace{-2.5pt} 3)}
\newcommand{\sdd}{\Sigma(360 \hspace{-2.5pt} \times \hspace{-2.5pt} 3)}
\newcommand{\llangle}{\langle\langle\,}
\newcommand{\rrangle}{\rangle\rangle\,}
\newcommand{\bone}{\mathbbm{1}}
\newcommand{\diag}{\mbox{diag}\,}
\newcommand{\zz}{\mathbbm{Z}}
\newcommand{\zzz}{\mathbbm{Z}_3}
\newcommand{\one}{\mathbf{1}}
\newcommand{\two}{\mathbf{2}}
\newcommand{\three}{\mathbf{3}}
\newcommand{\four}{\mathbf{4}}

\newcommand{\six}{\mathbf{6}}
\newcommand{\eight}{\mathbf{8}}
\newcommand{\nine}{\mathbf{9}}
\newcommand{\mc}{\mathcal{C}}

\allowdisplaybreaks

\swapnumbers 
\theoremstyle{definition} 
\newtheorem{define}{Definition}[section]

\theoremstyle{definition} 
\newtheorem{theorem}[define]{Theorem}

\begin{document}

\title{
\normalsize \hfill UWThPh-2010-10 \\[12mm]
\LARGE Principal series of finite subgroups of $SU(3)$
}

\author{
W.~Grimus\thanks{E-mail: walter.grimus@univie.ac.at}
\
and P.O.~Ludl\thanks{E-mail: patrick.ludl@univie.ac.at} 
\\*[3mm]
\small University of Vienna, Faculty of Physics \\
\small Boltzmanngasse 5, A--1090 Vienna, Austria
}

\date{29 July 2010}

\maketitle

\begin{abstract}
We attempt to give a complete description of the ``exceptional'' 
finite subgroups $\saa$, $\sbb$ and $\scc$ of $SU(3)$, with the aim to make
them amenable to model building for fermion masses and mixing.
The information on these groups which we derive contains 
conjugacy classes, proper normal subgroups, irreducible
representations, character tables and tensor products of their
three-dimensional irreducible representations. 
We show that, for these three exceptional groups, usage of their 
principal series, \textit{i.e.} ascending chains of normal subgroups,
greatly facilitates the computations and illuminates the relationship
between the groups. As a preparation and testing
ground for the usage of principal series, we study first 
the dihedral-like groups $\Delta(27)$ and $\Delta(54)$
because both are members of the principal series of 
the three groups discussed in the paper.
\end{abstract}

\newpage

\section{Introduction}
\label{introduction}

The masses of the fundamental fermions in the Standard Model are
spread over more than six orders of magnitude, if we consider the
charged fermions only. 
If we include neutrino masses, which are in the eV range or
smaller, we find the additional puzzle that neutrino masses must be
at least six orders of magnitude below the electron mass. 
Until today this mass problem is completely unsolved and no solution
is anywhere near the horizon. It may well be that it is easier to
understand fermion mixing than the fermion mass spectrum.
There are two mixing matrices, 
the Cabibbo--Kobayashi--Maskawa (CKM) matrix in the
quark sector and the Pontecorvo--Maki--Nakagawa--Sakata (PMNS) matrix
in the lepton sector. After the discovery that the Cabibbo angle is
approximately given by~\cite{gatto}
$\sin \theta_c \simeq \sqrt{m_d/m_s}$ where $m_d$ and $m_s$ are the
down and strange quark masses, respectively, there were many attempts
to relate mixing angles and quark masses by means of horizontal
symmetries---for early papers see for instance~\cite{ecker,wyler}. 
In the quark sector the CKM matrix is close to the unit matrix 
and, therefore, relations between the small mixing angles and a hierarchical
mass spectrum seems to be plausible. In the lepton sector, the
situation is completely different. The 
``tri-bimaximal'' mixing \textit{Ansatz}
\begin{equation}
U_\mathrm{PMNS} \simeq U_\mathrm{HPS} \equiv \left( \begin{array}{ccc}
2/\sqrt{6} & 1/\sqrt{3} & 0 \\ 
-1/\sqrt{6} & 1/\sqrt{3} & -1/\sqrt{2} \\ 
-1/\sqrt{6} & 1/\sqrt{3} & 1/\sqrt{2}
\end{array} \right)
\label{HPS}
\end{equation}
which has been put forward by Harrison,
Perkins and Scott (HPS)~\cite{HPS} is, 
at present, compatible with 
all the experimental data~\cite{results}.
In equation~(\ref{HPS}) we have left out possible Majorana phases.
The simple structure of the HPS mixing matrix rather suggests that
there is a symmetry responsible for lepton mixing and the
mixing angles are just numbers at the (high) scale where the symmetry is
effective, and corrections to $U_\mathrm{HPS}$ might be induced by
renormalization-group evolution of the mixing matrix down to the
electroweak scale.

Though it is by no means established that there is a symmetry origin
of quark and lepton mixing, it is nevertheless worthwhile to search for
possible candidates of symmetry groups. Since there are three families
of fundamental fermions, it is tempting to try $SU(3)$ and its
subgroups acting as family symmetries. Indeed, many models in the
literature are based upon finite non-abelian 
subgroups of $SU(3)$---see~\cite{review,tanimoto} for recent reviews
and~\cite{grimus} for a general discussion of extensions of the
Standard Model. If one uses abelian symmetries, one can, at best,
ensure that $\theta_{13} = 0$ in the mixing matrix while leaving the
other two mixing angles as free parameters~\cite{low}.

\begin{table}
\renewcommand{\arraystretch}{1.3}
\begin{center}
\begin{tabular}{|lr@{\hspace{5mm}}l|}
\hline 
subgroup & order & generators \\ \hline\hline
$\Sigma(60)$ & 60 & $A,\,E,\,W$ \\
$\Sigma(168)$ & 168 & $Y,\,E,\,Z$ \\ \hline
$\saa$ & 108 & $C,\,E,\,V$ \\
$\sbb$ & 216 & $C,\,E,\,V,\,X$ \\
$\scc$ & 648 & $C,\,E,\,V,\,D$ \\
$\sdd$ & 1080 & $A,\,E,\,W,\,F$\\ \hline
$\Delta(3n^2)$ ($n \geq 2$) & $3n^2$ & $A(n),\,E$ \\ 
$\Delta(6n^2)$ ($n \geq 1$) & $6n^2$ & $A(n),\,E,\,B$ \\ \hline
\end{tabular}
\end{center}
\caption{List of finite non-abelian subgroups of $SU(3)$ presented
  in~\cite{miller,fairbairn}. \label{subgrouplist}}
\end{table}
The finite subgroups of $SU(3)$ have been classified
nearly 100 years ago~\cite{miller}. Defining the constants 
\begin{equation}\label{constants}
\omega = e^{2\pi i/3}, \quad
\epsilon = e^{4\pi i/9}, \quad
\beta = e^{2\pi i/7}, \quad
\mu_1 = \frac{1}{2} \left( -1 + \sqrt{5} \right) ,\quad
\mu_2 = \frac{1}{2} \left( -1 - \sqrt{5} \right),
\end{equation}
we list the matrices 
\begin{subequations}
\label{generators}
\begin{eqnarray}
A(k) &=& \left( \begin{array}{ccc}
1 & 0 & 0 \\ 0 & e^{2\pi i/k} & 0 \\ 0 & 0 & e^{-2\pi i/k}
\end{array} \right) \quad ( k = 1,2,3,\ldots), 
\label{A} \\
B &=& \left( \begin{array}{ccc}
0 & 0 & -1 \\ 0 & -1 & 0 \\ -1 & 0 & 0
\end{array} \right), 
\label{B} \\
D &=& \left( \begin{array}{ccc}
\epsilon & 0 & 0 \\ 0 & \epsilon & 0 \\ 0 & 0 & \epsilon \omega 
\end{array} \right), 
\label{D} \\
E &=& \left( \begin{array}{ccc}
0 & 1 & 0 \\ 0 & 0 & 1 \\ 1 & 0 & 0
\end{array} \right), 
\label{E} \\
F &=& \left( \begin{array}{ccc}
-1 & 0 & 0 \\ 0 & 0 & -\omega \\ 0 & -\omega^2 & 0
\end{array} \right), \\
V &=& 
\frac{1}{\sqrt{3}\,i} \left( \begin{array}{ccc}
1 & 1 & 1 \\ 
1 & \omega & \omega^2 \\ 
1 & \omega^2 & \omega
\end{array} \right),
\label{V} \\
W &=& \frac{1}{2} \left( \begin{array}{ccc}
-1 & \mu_2 & \mu_1 \\ \mu_2 & \mu_1 & -1 \\ \mu_1 & -1 & \mu_2
\end{array} \right), \\
X &=& 
\frac{1}{\sqrt{3}\,i} \left( \begin{array}{ccc}
1 & 1 & \omega^2 \\ 
1 & \omega & \omega \\ 
\omega & 1 & \omega
\end{array} \right), 
\label{X} \\
Y &=& \left( \begin{array}{ccc}
\beta & 0 & 0 \\ 0 & \beta^2 & 0 \\ 0 & 0 & \beta^4 
\end{array} \right), \\
Z &=& 
\frac{i}{\sqrt{7}} \left( \begin{array}{ccc}
\beta^4 - \beta^3 & \beta^2 - \beta^5 & \beta   - \beta^6 \\ 
\beta^2 - \beta^5 & \beta   - \beta^6 & \beta^4 - \beta^3 \\ 
\beta   - \beta^6 & \beta^4 - \beta^3 & \beta^2 - \beta^5 
\end{array} \right),
\end{eqnarray}
\end{subequations}
which occur as group generators.
We use special symbols for 
\begin{equation}
A(2) \equiv A = \diag (1,\,-1,\,-1) 
\quad \mbox{and} \quad 
A(3) \equiv C = \diag (1,\, \omega,\, \omega^2).
\end{equation}
Then the list of finite subgroups of $SU(3)$, defined via their respective
sets of generators, is given in table~\ref{subgrouplist}.
This list consists of the two infinite series of ``dihedral-like''
subgroups denoted by the symbol $\Delta$ and six ``exceptional'' ones
denoted by $\Sigma$. 
It was noticed that table~\ref{subgrouplist} is not complete because
it does not contain all possible groups of types named (C) and (D)
in~\cite{miller}. However, the infinitely many groups missing in 
table~\ref{subgrouplist} are subgroups of either 
$\Delta(3n^2)$ (type (C)) or 
$\Delta(6n^2)$ (type (D))~\cite{bovier1,bovier,ludl,zwicky}.
An example is the Frobenius group $T_7$ with 21 elements
which is a subgroup of $\Delta(3 \!\times\! 7^2)$---for a discussion of
this group see~\cite{T7}.

Some of the groups in table~\ref{subgrouplist} are isomorphic to 
well-known alternating or symmetric groups:
\begin{equation}\label{iso}
\Delta(6) \simeq S_3, \quad \Delta(12) \simeq A_4, \quad 
\Delta(24) \simeq S_4, \quad \Sigma(60) \simeq A_5.
\end{equation}
Note that in the case of $\Delta(6)$ with $n=1$, 
the matrix $A(1)$ is identical with the unit matrix $\mathbbm{1}$ and
thus not a generator. 

Regarding the notation, we denoted the center of $SU(3)$ by 
\begin{equation}\label{center}
\zzz = \llangle \omega \bone \rrangle,
\end{equation}
where here and in the following 
$G = \llangle M_1, \ldots, M_r \rrangle$
means that $G$ is the group generated by the matrices 
$M_1, \ldots, M_r$.
As for the exceptional groups, we follow the notation of~\cite{fairbairn}
where for $n=36,\,72,\,216,\,360$ the groups denoted by 
$\Sigma(n\!\times\!3)$ have $3n$ elements and
contain the center of $SU(3)$, 
whereas the groups
\begin{equation}
\Sigma(n) \equiv \Sigma(n \!\times\! 3)/\zzz
\end{equation}
are \emph{not} subgroups of $SU(3)$; their character tables and
representations are discussed in~\cite{fairbairn}.
The groups $\Sigma(60)$ and $\Sigma(168)$ are simple~\cite{luhn1}, 
\textit{i.e.} they do not have any proper normal subgroups and,
therefore, cannot comprise the center of $SU(3)$. 
These groups were recently used for model building in~\cite{everett}
and \cite{king}, respectively.

The groups $S_3$, $A_4$ and $S_4$, 
according to equation~(\ref{iso}) dihedral-like groups with
$n=1$ and 2, are the most
favoured finite subgroups of $SU(3)$ for model building---see for
instance~\cite{review,tanimoto,S3,A4,pakvasa,S4} and references therein.
(A very early $A_4$ model for quarks is found in~\cite{wyler}.)
But also the groups $\Delta(27)$ and $\Delta(54)$ with
$n=3$ have been utilized for that purpose~\cite{Delta(27),Delta(54)}. 
In general, the dihedral-like groups, which are discussed in detail
in~\cite{bovier1,luhn2,escobar}, have the structure
\begin{equation}\label{dihdedral}
\Delta(3n^2) \cong (\zz_n \times \zz_n) \rtimes \zzz
\quad \mbox{and} \quad 
\Delta(6n^2) \cong (\zz_n \times \zz_n) \rtimes S_3,
\end{equation}
where the symbol ``$\rtimes$'' denotes the semidirect product defined
in appendix~\ref{semidirect}. 

In this paper we employ the concept of a 
\emph{principal series} (or chief series) 
of a group $G$~\cite{hall} in the context of finite
non-abelian subgroups of $SU(3)$.
In the following we use the symbol ``$\llhd$'' 
where the relation $H \llhd G$ indicates that $H$ is a proper normal
subgroup of $G$. A principal series of $G$ is defined by a series of
normal subgroups 
\begin{equation}\label{principal}
G_0 \equiv \{ e \} \llhd G_1 \llhd \cdots \llhd G_{\ell-1} \llhd G_\ell
\equiv G
\end{equation}
such that
\begin{enumerate}
\renewcommand{\labelenumi}{\roman{enumi}.}
\item
$G_j \llhd G_k$ holds for all $j < k$,
\item
$G_k/G_{k-1}$ is simple $\forall \: k=1, \ldots, \ell$.
\end{enumerate}
The latter property states that $G_{k-1}$ is a maximal normal subgroup
of $G_k$. Note that $G_k$ being a member of the principal series of
$G$ does not necessarily imply that a member of the principal
series of $G_k$ is part of the principal series of $G$. 
This will be exemplified by the
principal series discussed in this paper---compare 
equation~(\ref{csS72}) with equations~(\ref{csS216}) 
and~(\ref{csD54}).

If a principal series has a reasonable length $\ell$, 
it may be a useful concept 
\begin{itemize}
\item
to understand the structure of the group,
\item
to find the conjugacy classes, and
\item
to construct the irreducible representations (irreps) of $G$.
\end{itemize}
In order to elaborate on the last point we note that any irrep of a
factor group $G/G_k$ is also an irrep of $G$; 
a sensible investigation of the irreps which exploits the principal
series will start with $k=\ell$, which gives the trivial
one-dimensional irrep, and go on by descending to $k=0$,
\textit{i.e.} from the smallest to the largest group. Moreover, for
all indices $j,\,k$ with $0 \leq j < k \leq \ell$ the mapping 
\begin{equation}
h_{jk}: \; G/G_j \to G/G_k  \quad \mbox{with} \quad 
gG_j \mapsto gG_k
\end{equation}
is a homomorphism. Therefore, all irreps of $G/G_k$ with
$k=j+1,\ldots,\ell$ are also irreps of $G/G_j$.

Since the dihedral-like groups are covered in detail
by~\cite{bovier1,luhn2,escobar}, we concentrate---apart from
$\Delta(27)$ and $\Delta(54)$---on the exceptional
groups. As mentioned before, $\Sigma(60)$ and $\Sigma(168)$ are
simple, thus their principal series are trivial. Also the principal series 
$\{ e\} \llhd \zzz \llhd \sdd$, derived with the help of~\cite{gap}, is
too short for our purpose. 
However, the principal series
\begin{equation}\label{csS72}
\{ e \} \llhd \zzz \llhd \Delta(27) \llhd \Delta(54) \llhd \saa \llhd \sbb
\end{equation}
and
\begin{equation}\label{csS216}
\{ e \} \llhd \zzz \llhd \Delta(27) \llhd \Delta(54) \llhd \sbb \llhd \scc,
\end{equation}
which will be deduced from the group generators, look promising. 
The aim of the paper is a discussion 
of $\saa$, $\sbb$ and $\scc$ on the basis of 
equations~(\ref{csS72}) and~(\ref{csS216}).

The plan of the paper is as follows. Since both principal series above
contain the sequence $\Delta(27) \llhd \Delta(54)$, 
we start with the principal series 
\begin{equation}\label{csD54}
\{ e \} \llhd \zzz \llhd \zzz \times \zzz \llhd \Delta(27) 
\llhd \Delta(54) 
\end{equation}
in section~\ref{principal series D54}; there we also derive the
conjugacy classes and the character tables of $\Delta(27)$
and $\Delta(54)$, and finally we find all irreps of these two groups by
taking advantage of the principal series~(\ref{csD54}). 
Having done this exercise, we
apply our acquired knowledge to obtain the corresponding information
concerning the groups $\saa$, $\sbb$ and $\scc$ 
in sections~\ref{S36}, \ref{S72} and~\ref{S216}, respectively.
The conclusions are presented in section~\ref{concl}.

Some points are deferred to appendices: The definition and properties
of semidirect products of groups, which will often be used in this
paper, are expounded in appendix~\ref{semidirect}. A discussion of
tensor products of three-dimensional irreps of the dihedral-like
groups $\Delta(27)$ and $\Delta(54)$ is presented in appendix~\ref{delta}, 
the analogous but much more involved discussion of $\saa$ is developed in
appendix~\ref{tensor3}. Finally, in appendix~\ref{nine-216x3} 
we investigate the nine-dimensional irreps of $\scc$.

In checking the computations leading to the material presented in the
paper we have made extensive use of the computer algebra system
GAP~\cite{gap} which is very useful for the investigation of finite
groups. 

\section{The principal series of $\Delta(27)$ and $\Delta(54)$}
\label{principal series D54}
\subsection{The structure of $\Delta(27)$ and its irreducible representations}
\label{D27}
The three exceptional groups $\saa$, $\sbb$ and $\scc$, which we will
discuss in this paper, all contain the generators $C$ and $E$---see
table~\ref{subgrouplist}, therefore, 
\begin{equation}\label{ce}
\Delta (27) = \llangle C,\, E \rrangle
\end{equation}
is a subgroup of these three groups. 
We will later see that $\Delta (27)$ and also 
$\Delta(54)$ are even normal subgroups of $\saa$, $\sbb$ and $\scc$.

With the relations
\begin{equation}\label{conjCE}
C^{-1} E C = \omega E, \quad E^{-1} C E = \omega^2 C,
\end{equation}
the reordering relations
\begin{equation}\label{reordering}
EC = \omega CE, \quad
E^2 C = \omega^2 C E^2, \quad
E C^2 = \omega^2 C^2 E, \quad
E^2 C^2 = \omega C^2 E^2,
\end{equation}
the definition
\begin{equation}\label{cz}
\mathcal{Z} \equiv \{ \bone,\,  \omega \bone,\, \omega^2 \bone \}
\end{equation}
and the abbreviation 
$\mathcal{Z} \cdot g \equiv \{ g,\, \omega g,\, \omega^2 g \}$,
it is easy to write down the conjugacy classes:
\begin{equation}\label{cDelta27}
\begin{array}{llll}
\mc_1 = \{ \bone \},& \mc_2 = \{ \omega \bone \},&
\mc_3 = \{ \omega^2 \bone \}, & \\ 
\mc_4 = \mathcal{Z} \cdot C,  &  \mc_5 = \mathcal{Z} \cdot C^2,  &
\mc_6 = \mathcal{Z} \cdot E,  &  \mc_7 = \mathcal{Z} \cdot E^2,  \\ 
\mc_8 = \mathcal{Z} \cdot C E, &  \mc_9 = \mathcal{Z} \cdot C^2 E^2, & 
\mc_{10} = \mathcal{Z} \cdot C^2 E,  & \mc_{11} = \mathcal{Z} \cdot CE^2.
\end{array}
\end{equation}
Note that $\mathcal{Z}$ consists of 
the elements of the center~(\ref{center}) of $SU(3)$.
Furthermore, $\Delta (27)$ has the principal series
\begin{equation}\label{pDelta27}
\{ e \} \llhd \zzz \llhd \zzz \times \zzz \llhd \Delta(27)
\end{equation}
where $\zzz$ is given by equation~(\ref{center}), and 
\begin{equation}\label{3x3}
\zzz \times \zzz = 
\llangle \omega \bone \rrangle \times \llangle C \rrangle.
\end{equation}
The group $\zzz \times \zzz$ consists of the set of elements
\begin{equation}\label{H}
\mathcal{H} = \{ \mbox{diag} \left( \omega^a,\, \omega^b,\, \omega^{-a-b}
\right) | \, a,b = 0,1,2 \}.
\end{equation}

From equations~(\ref{ce}) and (\ref{3x3}) it follows that
\begin{equation}\label{fDelta27}
\Delta(27)/(\zzz \times \zzz) = \llangle \mathcal{H}E \rrangle,
\quad 
\Delta(27)/\zzz = \llangle \mathcal{Z}C,\, \mathcal{Z}E \rrangle.
\end{equation}
The cosets $\mathcal{Z}C$ and $\mathcal{Z}E$ commute, as can be read
off from the first relation in equation~(\ref{reordering}).
Therefore, the factor groups of the principal series~(\ref{pDelta27}) 
can be written as
\begin{equation}
\Delta(27)/(\zzz \times \zzz) \cong \zzz, \quad
\Delta(27)/\zzz \cong \zzz \times \zzz.
\end{equation}
As for the irreps of $\Delta (27)$, equation~(\ref{fDelta27})
immediately gives the one-dimensional irreps
\begin{equation}\label{1pq}
\one^{(p,q)}: \quad C \to \omega^p, \quad E \to \omega^q \quad
(p,q = 0,1,2).
\end{equation}
Note that, apart from the real irrep $\one^{(0,0)}$, all
one-dimensional irreps are pairwise complex conjugate to each
other, e.g., $\left( \one^{(1,2)} \right)^* = \one^{(2,1)}$. 
(This is a general feature of all complex one-dimensional irreps of any group.)
Since the number of conjugacy classes of $\Delta(27)$ is $n_c =
11$---see equation~(\ref{cDelta27})---and we have already found nine
irreps, there are two remaining ones. With the relation 
\begin{equation}\label{dim-ord}
\sum_{j=1}^{n_c} d_j^2 = \mbox{ord}\,(G),
\end{equation}
where the $d_j$ are the dimensions of the irreps and 
$\mbox{ord}\,(G)$ is the order of the group, \textit{i.e.} the number
of its elements, the remaining irreps have both dimension
three. Evidently they are given by
\begin{equation}
\begin{array}{ccc}
\three: & C \to C, & E \to E, \\
\three^*: & C \to C^*, & E \to E.
\end{array}
\end{equation}

\begin{table}
\begin{center}
\begin{tabular}{||c||ccccccccccc||}
\hline \hline
$\Delta(27)$ & $\mc_1$ & $\mc_2$ & $\mc_3$ & $\mc_4$ & $\mc_5$ &
$\mc_6$ & $\mc_7$ & $\mc_8$ & $\mc_9$ & $\mc_{10}$ & $\mc_{11}$ \\
(\# $C_k$) & (1) & (1) & (1) & (3) & (3) & (3) & (3) & (3) & (3) & (3) & (3) \\
$\mathrm{ord}(C_k)$ & 1 & 3 & 3 & 3 & 3 & 3 & 3 & 3 & 3 &  3 & 3 \\
\hline \hline
$\mathbf{1}^{(0,0)}$
& $1$ & $1$ & $1$ & $1$ & $1$ & $1$ & $1$ & $1$ & $1$ & $1$ & $1$ \\
$\mathbf{1}^{(0,1)}$
& $1$ & $1$ & $1$ & $1$ & $1$ & $\omega$ 
& $\omega^2$ & $\omega$ & $\omega^2$  & $\omega$ & $\omega^2$ \\ 
$\mathbf{1}^{(0,2)}$
& $1$ & $1$ & $1$ & $1$ & $1$ & $\omega^2$ 
& $\omega$ & $\omega^2$ & $\omega$  & $\omega^2$ & $\omega$ \\ 
$\mathbf{1}^{(1,0)}$
& $1$ & $1$ & $1$ & $\omega$ & $\omega^2$ & $1$ 
& $1$ & $\omega$ & $\omega^2$  & $\omega^2$ & $\omega$ \\ 
$\mathbf{1}^{(1,1)}$
& $1$ & $1$ & $1$ & $\omega$ & $\omega^2$ & $\omega$ & $\omega^2$ 
& $\omega^2$ & $\omega$  & $1$ & $1$ \\ 
$\mathbf{1}^{(1,2)}$
& $1$ & $1$ & $1$ & $\omega$ & $\omega^2$ & $\omega^2$ & $\omega$ 
& $1$ & $1$  & $\omega$ & $\omega^2$ \\ 
$\mathbf{1}^{(2,0)}$
& $1$ & $1$ & $1$ & $\omega^2$ & $\omega$ & $1$ & $1$ 
& $\omega^2$ & $\omega$  & $\omega$ & $\omega^2$ \\ 
$\mathbf{1}^{(2,1)}$
& $1$ & $1$ & $1$ & $\omega^2$ & $\omega$ & $\omega$ & $\omega^2$ 
& $1$ & $1$  & $\omega^2$ & $\omega$ \\ 
$\mathbf{1}^{(2,2)}$
& $1$ & $1$ & $1$ & $\omega^2$ & $\omega$ & $\omega^2$ & $\omega$ 
& $\omega$ & $\omega^2$& $1$ & $1$ \\ \hline 
$\three$ 
& $3$ & $3\omega$ & $3\omega^2$ 
& $0$ & $0$ & $0$ & $0$ & $0$ & $0$ & $0$ & $0$ \\
$\three^*$ 
& $3$ & $3\omega^2$ & $3\omega$ 
& $0$ & $0$ & $0$ & $0$ & $0$ & $0$ & $0$ & $0$ \\
\hline \hline
\end{tabular}
\caption{Character table of $\Delta(27)$.} \label{d27ct}
\end{center}
\end{table}
Now that we have found all classes and all irreps of $\Delta(27)$,
we can write down the character table---see
table~\ref{d27ct}. In that table, and also in all following character
tables, the second line indicates the number of elements in each
class, the third line the order of the elements in each class. The
order of an element $g$ of a group is defined as the smallest positive
power $m$ such that $g^m = e$ where $e$ is the unit element.

One can ask the question whether the two normal subgroups occurring
in the principal series~(\ref{pDelta27}) are \emph{all} normal subgroups of
$\Delta(27)$. To find all normal subgroups $N$ of a group $G$ one can apply
the following rules:
\begin{enumerate}
\item
The order of any subgroup $N$ is a divisor of the order of $G$.
\item
Any normal subgroup $N$ of $G$ must contain complete classes of $G$.
\end{enumerate}
Applying these rules to $\Delta(27)$, one quickly finds one additional normal
subgroup: 
\begin{equation}\label{N}
N = \llangle \omega \bone,\, E \rrangle.
\end{equation}
Thus, we can write down a second principal series
\begin{equation}\label{p1}
\{ e \} \llhd \zzz \llhd N \llhd \Delta(27).
\end{equation}
Obviously, $N$ is isomorphic to $\zzz \times \zzz$, which is formalized by 
$V^{-1} E V = C$ with $V$ given by equation~(\ref{V}). 
One can show by a simple computation that the principal series of
equation~(\ref{pDelta27}) can be obtained via a similarity
transformation with $V$ from the principal series~(\ref{p1}). 
This is a special case of the Jordan--H\"older
theorem~\cite{hall} which says that all 
principal series of a group $G$ are isomorphic.

\subsection{The group $\Delta(54)$}
\label{D54}

\paragraph{The conjugacy classes:}
This group is defined by 
\begin{equation}\label{def54}
\Delta (54) = \llangle C,\, E,\, B \rrangle = 
\llangle C,\, E,\, V^2 \rrangle.
\end{equation}
It is equivalent to use $B$ or $V^2$ because
\begin{equation}\label{V2}
B = E V^2 \quad \mbox{with} \quad
V^2 = \left( \begin{array}{ccc}
-1 & 0 & 0 \\ 0 & 0 & -1 \\ 0 & -1 & 0 
\end{array} \right). 
\end{equation}
The following theorems will be helpful for 
finding the conjugacy classes of $\Delta(54)$ and of the other groups
discussed in this paper.
\begin{theorem}\label{1}
Let $H$ be a proper normal subgroup of $G$ and $b \in G$ such that 
$b \not\in H$. If $C_k$ is a conjugacy class of $H$, then 
either $bC_kb^{-1} = C_k$ or 
the intersection between $bC_kb^{-1}$ and $C_k$ is empty.
\end{theorem}
\noindent
\textbf{Proof:} In order to prove this statement we will show that, if
there is an $h \in C_k$ such that 
$b h b^{-1} = h' \in C_k$, then this relation holds for \emph{all}
elements of $C_k$. The argument goes as follows. 
Since $C_k$ is a class of $H$, any element $\tilde h \in C_k$ can
be written as $\tilde h = h_1 h h_1^{-1}$ with $h_1 \in H$. Therefore, 
\begin{equation}\nonumber
b \tilde h b^{-1} = 
(b h_1 b^{-1}) (b h b^{-1}) (b h_1 b^{-1})^{-1} = 
(b h_1 b^{-1}) h' (b h_1 b^{-1})^{-1} \in C_k,
\end{equation}
because $b h_1 b^{-1} \in H$. Q.E.D.
\begin{theorem}\label{2}
Let $H$ be a proper normal subgroup of $G$ such that $G/H \cong \zz_r$ 
($r \geq 2$) and let $Hb$ be a generator of $G/H$. 
Then every conjugacy class of $G$ can be written in the form $S b^\nu$
where $S$ is a subset of $H$ and $\nu \in \{0,\,1\, \ldots,r-1 \}$. 
The conjugacy classes of $G$ which are subsets of $H$ can be obtained
from the conjugacy classes of $H$ in the following way:
\begin{enumerate}
\renewcommand{\labelenumi}{\roman{enumi}.}
\item
$C_k$ is a conjugacy class of $H$ such that $bC_kb^{-1} = C_k$. In
  this case $C_k$ is also a conjugacy class of $G$.
\item
$C_k$ is a conjugacy class of $H$ with empty 
intersection between $bC_kb^{-1}$ and $C_k$. Then the corresponding
conjugacy class of $G$ is obtained by 
$C_k \cup b C_k b^{-1} \cup \cdots \cup b^{r-1} C_k b^{-(r-1)}$.
\end{enumerate}
\end{theorem}
\noindent
\textbf{Proof:} With the above assumptions, 
the set of elements of $G$ can be written as a union of cosets:
\begin{equation}\nonumber
G = H \cup Hb \cup Hb^2 \cup \cdots \cup Hb^{r-1}.
\end{equation}
Note that $b^r \in H$ but $b^{r-1} \not\in H$. Obviously, 
$b (H b^\nu) b^{-1} = H b^\nu$ holds because $H$ is a normal subgroup of
$G$. For the same reason we deduce
\begin{equation}\nonumber
h (H b^\nu) h^{-1} = H (b^\nu h^{-1} b^{-\nu}) b^\nu = H b^\nu.
\end{equation}
Therefore, the cosets $H,\, Hb,\ldots,Hb^{r-1}$ are invariant under
$G$ and any class of $G$ must be contained in one of the cosets. This
proves the first part of the theorem. 
The second part is a consequence of theorem~\ref{1}. Q.E.D.

Now we observe $\Delta(54)/\Delta(27) \cong \zz_2$, which allows us to
use theorem~\ref{2} 
for constructing the conjugacy classes of $\Delta(54)$ from those of 
$\Delta(27)$. The matrix $B$ corresponds to the element $b$ occurring
in theorem~\ref{2}. We denote the classes of $\Delta(54)$ by $\mc'_k$.
With 
\begin{equation}\label{conjB}
B^{-1} C B = \omega^2 C^2, \quad B^{-1} E B = E^2
\end{equation}
we readily find 
\begin{equation}
\begin{array}{llll}
\mc'_1 = \mc_1, & \mc'_2 = \mc_2, & \mc'_3 = \mc_3, & \\
\mc'_4 = \mc_4 \cup \mc_5, &
\mc'_5 = \mc_6 \cup \mc_7, &
\mc'_6 = \mc_8 \cup \mc_9, &
\mc'_7 = \mc_{10} \cup \mc_{11},
\end{array}
\end{equation}
which exhausts the elements of $\Delta(27)$.
Then we first search for the class generated by $B$. The relations 
\begin{equation}
C^{-1} B C = \omega^2 CB, \quad E^{-1} B E = EB,
\end{equation}
together with equations~(\ref{conjCE}) and (\ref{reordering}) lead to 
\begin{eqnarray}
\mc'_8 &=& \{ B,\, EB,\, E^2B,\, CE^2B,\, C^2E^2B,\, \omega CEB,\,
\omega C^2B,\, \omega^2 CB,\, \omega^2 C^2EB \} \nonumber\\
&=& 
\{ V^2,\, EV^2,\, E^2V^2,\,  CV^2,\, C^2V^2,\, \omega CE^2V^2,\, 
\omega C^2EV^2,\, \omega^2 CEV^2,\, \omega^2 C^2E^2V^2 \}.
\nonumber \\ &&
\end{eqnarray}
It is then obvious that the missing conjugacy classes are given by
\begin{equation}
\mc'_9 = \omega C'_8, \quad \mc'_{10} = \omega^2 C'_8.
\end{equation}
Thus there are ten classes of $\Delta(54)$ and, therefore, this group
possesses ten inequivalent irreps.

Equations~(\ref{conjCE}) and (\ref{conjB}) establish the principal
series~(\ref{csD54}) of $\Delta(54)$. 
Concerning additional normal subgroups not included in the principal
series~(\ref{csD54}), the results developed at the end of
section~\ref{D27} apply for both $\Delta(27)$ and $\Delta(54)$, 
and the only additional proper normal subgroup is that of
equation~(\ref{N}). 

\paragraph{The factor groups:}
Now we want to discuss the factor groups of the principal
series~(\ref{csD54}); this will be useful for constructing the
irreps. First we have
\begin{equation}\label{fz2}
\Delta(54)/\Delta(27) \cong \zz_2.
\end{equation}
The generator of this $\zz_2$ is simply $\mathcal{D}B = \mathcal{D}V^2$,
if we denote by $\mathcal{D}$ the set of elements of $\Delta(27)$.

Using equation~(\ref{H}), we obtain
\begin{equation}\label{54/3x3}
\Delta(54)/(\zzz \times \zzz) = 
\llangle \mathcal{H}E,\, \mathcal{H}V^2 \rrangle.
\end{equation}
Remembering that the permutation group of three objects has the presentation
\begin{equation}
S_3: \quad a^2 = b^3 = (ab)^2 = e,
\end{equation}
we find that the assignment
\begin{equation}\label{ab}
a \to \mathcal{H}V^2, \quad b \to \mathcal{H}E
\end{equation}
constitutes an isomorphism because $(V^2)^2 = E^3 = (V^2 E)^2 = \bone$.
The $\zzz$ generated by $b$ is a normal subgroup of $S_3$ 
because $a b a^{-1} = aba = b^2$. Therefore, any
element of $S_3$ can be written in the form 
$b^m a^n$ with $m=0,1,2$ and $n=0,1$. In this way, $S_3$ has the
multiplication law
\begin{equation}
\left( b^{m_1} a^{n_1} \right) \left( b^{m_2} a^{n_2} \right) = 
b^{m_1} \left( a^{n_1} b^{m_2} a^{-n_1} \right) a^{(n_1+n_2)} =
b^{(m_1 + 2m_2n_1)} a^{(n_1+n_2)}
\end{equation}
and $S_3$ can be considered as the semidirect product 
$\zzz \rtimes \zz_2$. 
The definition of a semidirect product of groups and an overview of
its properties is given in appendix~\ref{semidirect}.
In summary we have proven that
\begin{equation}\label{fs3}
\Delta(54)/(\zzz \times \zzz) \cong S_3 \cong \zzz \rtimes \zz_2.
\end{equation}

Next we consider
\begin{equation}\label{54/3}
\Delta(54)/\zzz = 
\llangle \mathcal{Z}C,\, \mathcal{Z}E,\, \mathcal{Z}V^2 \rrangle.
\end{equation}
With $(V^2 C)^2 = \bone$, $CE = \omega^2 EC$ 
and the discussion related to $S_3$,
we notice that $\Delta(54)/\zzz$ has a structure similar to
$S_3$, but with $\zzz$ replaced by $\zzz \times \zzz$.
The assignment
\begin{equation}\label{abc}
a \to \mathcal{Z}V^2, \quad b \to \mathcal{Z}E, \quad
c \to \mathcal{Z}C
\end{equation}
allows to formulate the following presentation:
\begin{equation}
a^2 = b^3 = c^3 = (ab)^2 = (ac)^2 = e, \quad bc = cb.
\end{equation}
Therefore, we conclude
\begin{equation}\label{f18}
\Delta(54)/\zzz \cong (\zzz \times \zzz) \rtimes \zz_2.
\end{equation}
This group is one of the three non-abelian groups with 
18 elements, the other two being $D_9$ and $S_3 \times \zzz$~\cite{frampton}. 

\paragraph{The irreducible representations:}
Now we use the results of the factor groups to find all irreps of
$\Delta(54)$. Equation~(\ref{fz2}) immediately leads to the
two one-dimensional irreps
\begin{equation}\label{1-dim-irrep}
\begin{array}{cccc}
\one: & C \to 1, & E \to 1, & V^2 \to 1, \\
\one': & C \to 1, & E \to 1, & V^2 \to -1. 
\end{array}
\end{equation}
Next we search for irreps of $\Delta(54)$ by using equation~(\ref{fs3}). 
The one-dimensional irreps of $S_3$ are already taken care of by 
equation~(\ref{1-dim-irrep}). It remains to consider its
two-dimensional irrep which, by use of equation~(\ref{ab}), is
translated into an irrep of $\Delta(54)$. 
In a suitable basis, this irrep is given by the assignment 
\begin{equation}\label{2-dim-irrep}
\two: \quad C \to \left( \begin{array}{cc} 1 & 0 \\ 0 & 1 
\end{array} \right), \quad
E \to \left(
\begin{array}{cc} \omega & 0 \\ 0 & \omega^2
\end{array} \right), \quad 
\quad V^2 \to \left(
\begin{array}{cc} 0 & 1 \\ 1 & 0
\end{array} \right).
\end{equation}
Note that this irrep is real, with the representation matrix of $V^2$ acting
as the transformation matrix which switches between the irrep and its complex
conjugate. 
Moving to equation~(\ref{f18}), the inequivalent irreps of 
$(\zzz \times \zzz) \rtimes \zz_2$, which are not already contained in
equations~(\ref{1-dim-irrep}) and (\ref{2-dim-irrep}), are easily found with
our knowledge of $S_3$ and usage of equation~(\ref{abc}). 
There are three remaining inequivalent two-dimensional ones:
\begin{equation}\label{2-dim-irrep'}
\begin{array}{cccc}
\two': & 
C \to 
\left(
\begin{array}{cc} \omega & 0 \\ 0 & \omega^2
\end{array} \right), 
&
E \to 
\left( \begin{array}{cc} 1 & 0 \\ 0 & 1 
\end{array} \right), 
&
V^2 \to \left(
\begin{array}{cc} 0 & 1 \\ 1 & 0
\end{array} \right),
\\[4mm]
\two'': & 
C \to 
\left(
\begin{array}{cc} \omega & 0 \\ 0 & \omega^2
\end{array} \right), 
&
E \to 
\left( \begin{array}{cc} \omega & 0 \\ 0 & \omega^2
\end{array} \right), 
&
V^2 \to \left(
\begin{array}{cc} 0 & 1 \\ 1 & 0
\end{array} \right),
\\[4mm]
\two''': & 
C \to 
\left(
\begin{array}{cc} \omega & 0 \\ 0 & \omega^2
\end{array} \right), 
&
E \to 
\left( \begin{array}{cc} \omega^2 & 0 \\ 0 & \omega
\end{array} \right), 
&
V^2 \to \left(
\begin{array}{cc} 0 & 1 \\ 1 & 0
\end{array} \right).
\end{array}
\end{equation}
Since $\Delta(54)$ has ten classes, four irreps are missing. They are easy to
guess because the definition~(\ref{def54}) itself constitutes a
three-dimensional irrep, out of which we find three other ones by complex
conjugation and by multiplication with $\one'$ of
equation~(\ref{1-dim-irrep}): 
\begin{equation}
\begin{array}{cccc}
\three: & C \to C, & E \to E, & V^2 \to V^2, \\
(\three)^*: & C \to C^*, & E \to E, & V^2 \to V^2, \\
\three': & C \to C, & E \to E, & V^2 \to -V^2, \\
(\three')^*: & C \to C^*, & E \to E, & V^2 \to -V^2.
\end{array}
\end{equation}
\begin{table}
\begin{center}
\begin{tabular}{||c||cccccccccc||}
\hline \hline
$\Delta(54)$ & $\mc'_1$ & $\mc'_2$ & $\mc'_3$ & $\mc'_4$ & $\mc'_5$ &
$\mc'_6$ & $\mc'_7$ & $\mc'_8$ & $\mc'_9$ & $\mc'_{10}$ \\
(\# $C_k$) & (1) & (1) & (1) & (6) & (6) & (6) & (6) & (9) & (9) & (9) \\
$\mathrm{ord}(C_k)$ & 1 & 3 & 3 & 3 & 3 & 3 & 3 & 2 & 6 & 6 \\
\hline \hline
$\one$
& $1$ & $1$ & $1$ & $1$ & $1$ & $1$ & $1$ & $1$ & $1$ & $1$ \\
$\one'$
& $1$ & $1$ & $1$ & $1$ & $1$ & $1$ 
& $1$ & $-1$ & $-1$ & $-1$ \\ \hline
$\two$
& $2$ & $2$ & $2$ & $2$ & $-1$ 
& $-1$ & $-1$ & $0$ & $0$ & $0$ \\ 
$\two'$
& $2$ & $2$ & $2$ & $-1$ & $2$ 
& $-1$ & $-1$ & $0$ & $0$ & $0$ \\ 
$\two''$
& $2$ & $2$ & $2$ & $-1$ & $-1$ 
& $-1$ & $2$ & $0$ & $0$ & $0$ \\ 
$\two'''$
& $2$ & $2$ & $2$ & $-1$ & $-1$ 
& $2$ & $-1$ & $0$ & $0$ & $0$ \\ \hline 
$\three$ 
& $3$ & $3\omega$ & $3\omega^2$ 
& $0$ & $0$ & $0$ & $0$ & $-1$ & $-\omega$ & $-\omega^2$ \\
$\three'$ 
& $3$ & $3\omega$ & $3\omega^2$ 
& $0$ & $0$ & $0$ & $0$ & $1$ & $\omega$ & $\omega^2$ \\
$\three^*$ 
& $3$ & $3\omega^2$ & $3\omega$ 
& $0$ & $0$ & $0$ & $0$ & $-1$ & $-\omega^2$ & $-\omega$ \\
$\left( \three' \right)^*$ 
& $3$ & $3\omega^2$ & $3\omega$ 
& $0$ & $0$ & $0$ & $0$ & $1$ & $\omega^2$ & $\omega$ \\
\hline \hline
\end{tabular}
\caption{Character table of $\Delta(54)$.} \label{d54ct}
\end{center}
\end{table}

The irreps we have constructed here agree with those given
in~\cite{escobar}. The character table of $\Delta(54)$ is presented in
table~\ref{d54ct}. 
This concludes the discussion of the irreps of $\Delta(27)$
and $\Delta (54)$ in the light of their principal series.

\section{The group $\saa$}
\label{S36}

\subsection{Definition and principal series}
This group is generated by
\begin{equation}\label{def}
\saa = \langle\langle\, C,\,E,\, V\, \rangle\rangle.
\end{equation}
It is easy to check that the generators of the group fulfill
\begin{equation}\label{presentation108}
C^3 = E^3 = V^4 = \bone, \quad
E V^{-1} C V = E V C^{-1} V^{-1} = (EC)^3 = \bone.
\end{equation}
According to GAP~\cite{gap}, these relations establish
a presentation of $\saa$. 

For the following considerations we need the set of relations 
\begin{equation}\label{ECconjV}
C^{-1} V C = C^2 E V, \quad  E^{-1} VE = \omega C^2 E^2 V
\end{equation}
and
\begin{equation}\label{conjV}
V^{-1} C V = E^2, \quad \quad V^{-1} E V = C
\end{equation}
With equations~(\ref{V2}) and (\ref{conjV}) 
and taking into account the definitions of
$\zzz$, $\Delta(27)$ and $\Delta(54)$, one evidently finds the principal
series  
\begin{equation}\label{ps36x3}
\{ e \} \llhd \zzz \llhd \Delta(27) \llhd \Delta(54) \llhd \saa.
\end{equation}
Note that $\zzz \times \zzz$ of equation~(\ref{3x3}) is not part of
this principal series because it is not invariant under $V$---see
equation~(\ref{conjV}). With the rules expounded at the end of
section~(\ref{D27}) one can show that the principal series~(\ref{ps36x3})
comprises all normal subgroups of~$\saa$.

\subsection{Conjugacy classes}
\label{conjugacy classes 36x3}

The reordering relations~(\ref{reordering}) and 
\begin{equation}\label{reordV}
VC = EV, \quad VE = C^2V
\end{equation}
allow to write every element of $\saa$ in the form 
\begin{equation}\label{standardform}
\omega^p C^q E^r V^s \quad \mbox{with} \quad
p,q,r = 0,1,2, \quad s = 0,1,2,3.
\end{equation}
In order to find the classes of the group we will take advantage of
this fact. Moreover, since $\saa/\Delta(54) \cong \zz_2$, we can make
use of theorem~\ref{2}.
Then, with equations~(\ref{conjCE}), (\ref{ECconjV}) and
(\ref{conjV}), it is tedious but 
straightforward to find all the classes. 
We use the principal series~(\ref{ps36x3}) 
as a guideline for ordering the classes.

Each element of the center of the group is a class of its own,
\textit{i.e.}
\begin{equation}
C_1 = \left\{ \bone \right\}, \quad
C_2 = \left\{ \omega \bone \right\}, \quad
C_3 = \left\{ \omega^2 \bone \right\}.
\end{equation}
Next we turn to the elements $C$ and $E$ which generate
$\Delta(27)$. It turns out that these two elements generate the same
class of $\saa$, 
whereas the product $CE$ generates a different class:
\begin{eqnarray}
C_4 & = & \left\{
C,\, \omega C,\, \omega^2 C,\, C^2,\, \omega C^2,\, \omega^2 C^2,\,
E,\, \omega E,\, \omega^2 E,\, E^2,\, \omega E^2,\, \omega^2 E^2
\right\},
\\ 
C_5 & = & \left\{
CE,\, \omega CE,\, \omega^2 CE,\, 
C^2 E,\, \omega C^2 E,\, \omega^2 C^2 E, 
\right. \nonumber  \\ && \left. \:
C E^2,\, \omega C E^2,\, \omega^2 C E^2,\, 
C^2 E^2,\, \omega C^2 E^2,\, \omega^2 C^2 E^2
\right\}.
\end{eqnarray}
These two classes have 12 elements each and together with $C_{1,2,3}$
they exhaust the subgroup $\Delta(27)$. Note that $C_4$ and $C_5$ are
invariant under multiplication of their elements by $\omega$, 
\textit{i.e.} symbolically $\omega C_k = C_k$ for $k=4,5$.
Now we go on to $\Delta(54)$ and consider the class generated by $V^2$: 
\begin{equation}
C_6 = \left\{
V^2,\, EV^2,\, E^2V^2,\, CV^2,\, C^2V^2,\,
\omega CE^2V^2,\, \omega C^2EV^2,\, \omega^2 CEV^2,\, \omega^2 C^2E^2V^2
\right\}. 
\end{equation}
This class is identical with $\mc'_8$ of $\Delta(54)$.
Obviously, in this case multiplication of the elements of $C_6$ by
powers of $\omega$ gives two new classes:
\begin{equation}
C_7 = \omega C_6, \quad C_8 = \omega^2 C_6.
\end{equation}
Eventually, we are left with the elements of $\saa$ which are not
contained in $\Delta(54)$. The class generated by $V$ is obtained as
\begin{equation}
C_9 = \left\{
V,\, \omega CV,\, \omega C^2V,\,
\omega EV,\, \omega E^2V,\,
C^2EV,\, CE^2V,\, \omega CEV,\, \omega C^2E^2V \right\},
\end{equation}
which immediately results in the two further classes
\begin{equation}
C_{10} = \omega C_9, \quad C_{11} = \omega^2 C_9.
\end{equation}
Finally, we are left with 
\begin{equation}
C_{12} = \left\{
V^3,\, \omega^2 CV^3,\, \omega^2 C^2V^3,\,
\omega^2 EV^3,\, \omega^2 E^2V^3,\,
\omega^2 C^2EV^3,\, \omega^2 CE^2V^3,\, CEV^3,\, C^2E^2V^3 \right\}
\end{equation}
and
\begin{equation}
C_{13} = \omega C_{12}, \quad C_{14} = \omega^2 C_{12}.
\end{equation}
Noting that $V^3 = V^\dagger = V^*$, we see that $C_{12}$ can be
obtained from $C_9$ by complex conjugation.

\subsection{Factor groups and irreps}
\label{factor groups and irreps}

Finding the factor groups $G/G_k$---see introduction---with respect to
the principal series~(\ref{ps36x3}) is almost trivial:
\begin{equation}\label{f36x3}
\saa/\Delta(54) \cong \zz_2, \quad 
\saa/\Delta(27) \cong \zz_4, \quad 
\saa/\zzz \equiv \Sigma(36).
\end{equation}
The first two relations follow from the property of the generator $V$.
As discussed in the introduction, the last relation is simply the
definition of $\Sigma(36)$. 

From the first two factors in equation~(\ref{f36x3}) 
it follows immediately that there are four one-dimensional irreps,
characterized by
\begin{equation}\label{1-dim-36x3}
\one^{(p)}: \quad C \to 1,\quad E \to 1,\; V \to i^p
\quad (p = 0,1,2,3).
\end{equation}

Next we move to the group $\Sigma(36)$ whose elements consist 
of the cosets $g\mathcal{Z}$, where 
$\mathcal{Z}$ defined in equation~(\ref{cz})
contains the elements of the center of $\saa$ and $g \in \saa$.
Using the results of subsection~\ref{conjugacy classes 36x3}, it is
straightforward to find the classes $\widetilde C_k$ of $\Sigma(36)$
from those of $\saa$:
\begin{equation}
\begin{array}{ccl}
\widetilde C_1 &=& \mathcal{Z}, \\
\widetilde C_2 &=& C_4 \mathcal{Z}, \\
\widetilde C_3 &=& C_5 \mathcal{Z}, \\
\widetilde C_4 &=& C_6 \mathcal{Z} \equiv C_7 \mathcal{Z} \equiv 
C_8 \mathcal{Z}, \\
\widetilde C_5 &=& C_9 \mathcal{Z} \equiv C_{10} \mathcal{Z} \equiv 
C_{11} \mathcal{Z}, \\
\widetilde C_6 &=& C_{12} \mathcal{Z} \equiv C_{13} \mathcal{Z} \equiv 
C_{14} \mathcal{Z}.
\end{array}
\end{equation}
Therefore, $\Sigma(36)$ possesses six irreps. These are also 
irreps of $\saa$ such that the center of $\saa$ is mapped onto unity.
This is the case with the four one-dimensional irreps of
equation~(\ref{1-dim-36x3}), therefore, we already know four of the
six irreps of $\Sigma(36)$. Denoting the dimensions of the remaining
two irreps of $\Sigma(36)$ by $d$ and $d'$, equation~(\ref{dim-ord})
tells us that
\begin{equation}
4 \times 1^2 + d^2 + {d'}^2 = 36.
\end{equation}
This equation has a single solution for the two dimensions, namely
\begin{equation}
d = d' = 4.
\end{equation}
We denote the corresponding irreps by $\four$ and $\four^\prime$.

Since $E^{-1} C^{-1} EC = \omega \bone \in \zzz$ which is
mapped onto unity for $\Sigma(36)$,
we find that $C$ and $E$ commute in the four-dimensional
irreps. Therefore, in these irreps we can adopt a basis where both $C$
and $E$ are diagonal. Equation~(\ref{conjV}) gives 
$V^{-2} C V^2 = C^2$ and $V^{-2} E V^2 = E^2$. Thus $C$
and $C^2$ are both diagonal and have the same eigenvalues. 
Since $C^3 = \bone$, the eigenvalues must be 
powers of $\omega$. Up to reorderings there are two solutions for $C$,
namely 
\begin{equation}
C \to \diag(1,\omega,1,\omega^2)
\quad \mbox{and} \quad
\diag(\omega,\omega,\omega^2,\omega^2).
\end{equation}
The same applies to the representation of $E$, only the ordering must
be different in order to ensure that $C$ and $E$ are represented
differently. A method for finding explicit realizations of the
four-dimensional irreps is presented in appendix~\ref{tensor3}. 
The result is given by
\begin{equation}\label{four}
\begin{array}{cccc}
\four: & C \to \diag (1,\omega,1,\omega^2), & 
E \to \diag (\omega,1,\omega^2,1), & 
V \to \left( 
\begin{array}{cccc}
0 & 1 & 0 & 0 \\ 
0 & 0 & 1 & 0 \\ 
0 & 0 & 0 & 1 \\ 
1 & 0 & 0 & 0 
\end{array} \right)
\end{array}
\end{equation}
and
\begin{equation}\label{four'}
\begin{array}{cccc}
\four^\prime: & C \to \diag (\omega,\omega,\omega^2,\omega^2), & 
E \to \diag (\omega,\omega^2,\omega^2,\omega), & 
V \to \left( 
\begin{array}{cccc}
0 & 1 & 0 & 0 \\ 
0 & 0 & 1 & 0 \\ 
0 & 0 & 0 & 1 \\ 
1 & 0 & 0 & 0 
\end{array} \right).
\end{array}
\end{equation}
One can check, for instance, that the
relations~(\ref{presentation108}) are fulfilled. 
\begin{table}
\begin{center}
\begin{tabular}{||c||cccccc||}
\hline \hline
$\Sigma(36)$\rule{0mm}{15pt} & 
$\widetilde C_1$ & $\widetilde C_2$ & $\widetilde C_3$ &
$\widetilde C_4$ & $\widetilde C_5$ & $\widetilde C_6$ \\ 
(\# $C_k$) & (1) & (4) & (4) & (9) & (9) & (9) \\
$\mathrm{ord}(C_k)$ & 1 & 3 & 3 & 2 &  4 & 4\\
\hline \hline
$\mathbf{1}^{(1)}$
& $1$ & $1$ & $1$ & $1$ & $1$ & $1$ \\
$\mathbf{1}^{(2)}$
& $1$ & $1$ & $1$ & $-1$ & $i$ & $-i$ \\
$\mathbf{1}^{(3)}$
& $1$ & $1$ & $1$ & $1$ & $-1$ & $-1$ \\
$\mathbf{1}^{(4)}$
& $1$ & $1$ & $1$ & $-1$ & $-i$ & $i$ \\
\hline
$\mathbf{4}$
& $4$ & $1$ & $-2$ & $0$ & $0$ & $0$ \\
$\mathbf{4}^\prime$
& $4$ & $-2$ & $1$ & $0$ & $0$ & $0$ \\
\hline \hline
\end{tabular}
\caption{Character table of $\Sigma(36)$.}\label{s36ct}
\end{center}
\end{table}

\begin{table}
\begin{center}
\rotatebox{90}{
\begin{tabular}{||c||ccc|c|c|ccc|ccc|ccc||}
\hline \hline
$\saa$ & 
$C_1$ & $C_2$ & $C_3$ & $C_4$ & $C_5$ & $C_6$ & $C_7$ & $C_8$ &
$C_9$ & $C_{10}$ & $C_{11}$ & $C_{12}$ & $C_{13}$ & $C_{14}$ \\
(\# $C_k$) & (1) & (1) & (1) & (12) & (12) & (9) & (9) & (9) &
(9) & (9) & (9) & (9) & (9) & (9) \\
$\mathrm{ord}(C_k)$ & 1 & 3 & 3 & 3 & 3 & 2 & 6 & 6 & 4 & 12 & 12 & 4
& 12 & 12 \\ 
\hline \hline
$\one^{(0)}$ & $1$ & $1$ & $1$ & $1$ & $1$ 
& $1$ & $1$ & $1$         
& $1$ & $1$ & $1$         
& $1$ & $1$ & $1$ \\      
$\one^{(1)}$ & $1$ & $1$ & $1$ & $1$ & $1$ 
& $-1$ & $-1$ & $-1$
& $i$ & $i$ & $i$ 
& $-i$ & $-i$ & $-i$ \\
$\one^{(2)}$ & $1$ & $1$ & $1$ & $1$ & $1$ 
& $1$ & $1$ & $1$
& $-1$ & $-1$ & $-1$ 
& $-1$ & $-1$ & $-1$ \\
$\one^{(3)}$ & $1$ & $1$ & $1$ & $1$ & $1$ 
& $-1$ & $-1$ & $-1$
& $-i$ & $-i$ & $-i$ & 
$i$ & $i$ & $i$ \\
\hline
$\three^{(0)}$ & $3$ & $3 \omega$ & $3 \omega^2$ &
$0$ & $0$ &
$-1$ & $-\omega$ & $-\omega^2$ &
$1$ & $\omega$ & $\omega^2$ &
$1$ & $\omega$ & $\omega^2$ \\
$\three^{(1)}$ & $3$ & $3 \omega$ & $3 \omega^2$ &
$0$ & $0$ &
$1$ & $\omega$ & $\omega^2$ &
$i$ & $i \omega$ & $i \omega^2$ &
$-i$ & $-i \omega$ & $-i \omega^2$ \\
$\three^{(2)}$ & $3$ & $3 \omega$ & $3 \omega^2$ &
$0$ & $0$ &
$-1$ & $-\omega$ & $-\omega^2$ &
$-1$ & $-\omega$ & $-\omega^2$ &
$-1$ & $-\omega$ & $-\omega^2$ \\
$\three^{(3)}$ & $3$ & $3 \omega$ & $3 \omega^2$ &
$0$ & $0$ &
$1$ & $\omega$ & $\omega^2$ &
$-i$ & $-i \omega$ & $-i \omega^2$ &
$i$ & $i \omega$ & $i \omega^2$ \\
$\left( \three^{(0)} \right)^*$ & $3$ & $3 \omega^2$ & $3 \omega$ &
$0$ & $0$ &
$-1$ & $-\omega^2$ & $-\omega$ &
$1$ & $\omega^2$ & $\omega$ &
$1$ & $\omega^2$ & $\omega$ \\
$\left( \three^{(1)} \right)^*$ & $3$ & $3 \omega^2$ & $3 \omega$ &
$0$ & $0$ &
$1$ & $\omega^2$ & $\omega$ &
$-i$ & $-i \omega^2$ & $-i \omega$ &
$i$ & $i \omega^2$ & $i \omega$ \\
$\left( \three^{(2)} \right)^*$ & $3$ & $3 \omega^2$ & $3 \omega$ &
$0$ & $0$ &
$-1$ & $-\omega^2$ & $-\omega$ &
$-1$ & $-\omega^2$ & $-\omega$ &
$-1$ & $-\omega^2$ & $-\omega$ \\
$\left( \three^{(3)} \right)^*$ & $3$ & $3 \omega^2$ & $3 \omega$ &
$0$ & $0$ &
$1$ & $\omega^2$ & $\omega$ &
$i$ & $i \omega^2$ & $i \omega$ &
$-i$ & $-i \omega^2$ & $-i \omega$ \\
\hline
$\mathbf{4}$ & $4$ & $4$ & $4$ &
$1$ & $-2$ &
$0$ & $0$ & $0$ &
$0$ & $0$ & $0$ &
$0$ & $0$ & $0$ \\
$\mathbf{4}^\prime$ & $4$ & $4$ & $4$ &
$-2$ & $1$ &
$0$ & $0$ & $0$ &
$0$ & $0$ & $0$ &
$0$ & $0$ & $0$ \\
\hline \hline
\end{tabular}
}
\caption{Character table of $\saa$. 
\label{s36x3ct}}
\end{center}
\end{table}
Up to now we have found six irreps of $\saa$. Since this group possesses 14
classes, there are eight remaining irreps. We guess that these have
all dimensions three because then 
\begin{equation}
4 \times 1^2 + 2 \times 4^2 + 8 \times 3^2 = 108.
\end{equation}
This is indeed easily proven.
Noting that the only $3 \times 3$ matrix which commutes with
both $C$ and $E$ is proportional to the unit matrix, \textit{i.e.} $C$
and $E$ generate a three-dimensional irrep of $\Delta(27)$, it is
obvious that the defining representation~(\ref{def}) of $\saa$
is irreducible as well. With these considerations we use the
one-dimensional irreps and complex conjugation to construct
the eight three-dimensional irreps:
\begin{equation}\label{3dimirreps}
\begin{array}{cccc}
\three^{(p)}: & C \to C, & E \to E, & V \to i^p V, \\
\left( \three^{(p)} \right)^*: 
& C \to C^*, & E \to E, & V \to (-i)^p V^*, 
\end{array}
\end{equation}
where $p = 0,1,2,3$. Therefore, 
$\three^{(p)} \equiv \one^{(p)} \otimes \three^{(0)}$.
The character table of $\saa$ is given in table~\ref{s36x3ct}.
The vertical lines in this table indicate the classes which merge into
one class when one makes the transition from $\saa$ to $\Sigma(36)$.

\section{The group $\sbb$}
\label{S72}

\subsection{Definition and principal series}
\label{subsbb}

The group is generated by
\begin{equation}
\sbb = \llangle C,\, E,\, V,\, X\,\rrangle.
\end{equation} 
Obviously, $\saa$ is a subgroup of $\sbb$. 
Let us investigate the properties of the new generator $X$. 
First of all we find 
\begin{equation}\label{ordX}
X^2 = C^2 V^2 \in \saa, \quad X^4 = \bone.
\end{equation}
For the computation of the classes of $\sbb$ we need 
\begin{equation}\label{conjX}
XCX^{-1} = C^2 E, \quad
XEX^{-1} = \omega^2 CE, \quad
XVX^{-1} = \omega^2 EV^3
\end{equation}
and
\begin{equation}
C^{-1} X C = CEX, \quad E^{-1} X E = \omega CX, \quad 
V^{-1} X V = \omega^2 CV^2X.
\end{equation}
From equation~(\ref{conjX}) we read off that the principal series of
$\sbb$ is given by~equation~(\ref{csS72}).

One can ask the question if the four normal subgroups occurring in the
principal series~(\ref{csS72}) comprise all normal subgroups of
$\sbb$. The answer is no. 
Laborious application of the procedure outlined at the end of
section~\ref{D27} yields one more normal subgroup given by 
\begin{equation}
N = \llangle C,\, E,\, X \rrangle.
\end{equation}
Performing a similarity transformation with the matrix $D$ of
equation~(\ref{D}), we obtain
\begin{equation}\label{simtraf}
D^{-1} C D = C, \quad D^{-1} E D = CE, \quad D^{-1} X D = V,
\end{equation}
which demonstrates that $N$ is isomorphic to $\saa$.
If we replace in the principal series~(\ref{csS72}) the subgroup $\saa$ by
$N$, we obtain an isomorphic principal series with the isomorphism given by the
transformation~(\ref{simtraf}); this is again a
manifestation of the Jordan--H\"older theorem.

\subsection{Conjugacy classes} 

With the reordering relations
\begin{equation}\label{reordX}
XC = C^2 EX, \quad 
XE = \omega^2 CEX, \quad
XV = \omega^2 EV^3 X.
\end{equation}
and with equation~(\ref{ordX}) we 
conclude that every element of $\sbb$ can be written as
\begin{equation}\label{standardform72}
\omega^p C^q E^r V^s X^t \quad \mbox{with} \quad
p,q,r = 0,1,2, \quad s = 0,1,2,3, \quad t = 0,1.
\end{equation}
With theorem~\ref{2} and equation~(\ref{conjX}) 
we first search for the classes $C'_k$ of $\sbb$ which are subsets of $\saa$. 
Going through the classes $C_j$ 
$(j = 1,...,14)$ of $\saa$, one derives,
with some patience, the upper part of table~\ref{classes72x3}.
\begin{table}
	\begin{center}
	\begin{tabular}{|r@{\,=\,}l|r|}
	\hline
\multicolumn{2}{|c}{Classes \hphantom{xxxx}} & 
Number of elements \\
	\hline\hline
\hspace{10mm} $C_1'$ & $C_1 = C_\bone$ & 1\hspace{15mm} \\ 
	$C_2'$ & $\omega C'_1$ & 1\hspace{15mm} \\ 
	$C_3'$ & $\omega^2 C'_1$ & 1\hspace{15mm} \\
	$C_4'$ & $C_4\cup C_5 =C_C$ & 24\hspace{15mm} \\
	$C_5'$ & $C_6 = C_{V^2}$ & 9\hspace{15mm} \\
	$C_6'$ & $\omega C'_5$ & 9\hspace{15mm} \\
	$C_7'$ & $\omega^2 C'_5$ & 9\hspace{15mm} \\
	$C_8'$ & $C_9\cup C_{12} =C_V$ & 18\hspace{15mm} \\
	$C_9'$ & $\omega C_8'$ & 18\hspace{15mm} \\
	$C_{10}'$ & $\omega^2 C_8'$ & 18\hspace{15mm} \\
	\hline
$C'_{11}$ & $C_X$ & 18\hspace{15mm} \\ 
$C'_{12}$ & $\omega C_{X}$ & 18\hspace{15mm} \\ 
$C'_{13}$ & $\omega^2 C_{X}$ & 18\hspace{15mm} \\
$C'_{14}$ & $C_{VX}$ & 18\hspace{15mm} \\
$C'_{15}$ & $\omega C_{VX}$ & 18\hspace{15mm} \\
$C'_{16}$ & $\omega^2 C_{VX}$ & 18\hspace{15mm} \\
\hline
\multicolumn{2}{|c}{Total number of elements} & 216\hspace{15mm} \\
	\hline
	\end{tabular}
	\end{center}
\caption{The conjugacy classes of $\sbb$. \label{classes72x3}}
\end{table}
The remaining classes must have elements with the standard form 
$\omega^p C^q E^r V^s X$. Let us start with the class which contains $X$:
	\begin{equation}
	\begin{split}
	C_{X}=\{
	& X,\,
	\omega^2 V^2 X,\,
	\omega^2 E^2 V^2 X,\,
	\omega^2 C^2 E V^2 X,\,
	\omega C^2 X,\,
	\omega E V^2 X,\,
	C^2 V^2 X,\,
	\omega C X,\, \\
	& \omega^2 C V^2 X,\,
	\omega^2 C E^2 V^2 X,\,
	\omega E X,\,
	\omega C E V^2 X,\,
	\omega C^2 E X,\,
	C^2 E^2 V^2 X,\,
	C E^2 X,\, \\
	& \omega^2 C^2 E^2 X,\,
	C E X,\,
	E^2 X\}.
	\end{split}
	\end{equation}
Since $\omega X,\omega^2 X\not\in C_X$, we obtain the next two classes as
	\begin{equation}
	C_{\omega X}=\omega C_X,\quad C_{\omega^2 X}=\omega^2 C_X.
	\end{equation}
The simplest element that is not contained in the classes already listed is
$VX$. Its associated class is given by
	\begin{equation}
	\begin{split}
	C_{VX}=\{
	& V X,\,
	C V^3 X,\,
	\omega C E V^3 X,\,
	\omega^2 C^2 E V^3 X,\,
	\omega^2 E V X,\,
	\omega^2 C^2 V X,\,
	C E^2 V^3 X,\, \\
	& \omega C^2 E^2 V^3 X,\,
	\omega^2 C E V X,\,
	\omega C E^2 V X,\,
	\omega E^2 V^3 X,\,
	\omega^2 C V X,\,
	C^2 E V X,\, \\
	& E^2 V X,\,
	\omega C^2 V^3 X,\,
	\omega V^3 X,\,
	C^2 E^2 V X,\,
	\omega^2 E V^3 X\}.
	\end{split}
	\end{equation}
Since $\omega VX,\omega^2 VX\not\in C_{VX}$, the remaining two classes are
\begin{equation}
C_{\omega VX}=\omega C_{VX},\quad C_{\omega^2 VX}=\omega^2 C_{VX}.
\end{equation}
The conjugacy classes we have found are listed in
table~\ref{classes72x3}, from 
where we also read off that we have accommodated all elements of
$\sbb$. Therefore, the set of classes listed in table~\ref{classes72x3} 
is complete. We have also indicated one member of each class.
Now we face the task of finding all 16 irreps of $\sbb$,
which will be facilitated by finding first the factor groups.

\subsection{The factor groups}

The factor groups corresponding to the principal series~(\ref{csS72})
are given by
\begin{equation}\label{f72x3}
\sbb/\saa \cong \zz_2, \quad
\sbb/\Delta(54) \cong \zz_2 \times \zz_2, \quad
\sbb/\Delta(27) \cong Q_8.
\end{equation}
In addition, there is the factor group
$\sbb/\zzz \equiv \Sigma(72)$. 
The first statement of equation~(\ref{f72x3}) follows from
equation~(\ref{ordX}). 
In order to prove the second statement, we denote the set of matrices
of $\Delta(54)$ by $\bar{\mathcal{D}}$. Then, 
\begin{equation}\label{sbb54}
\sbb/\Delta(54) = 
\llangle \bar{\mathcal{D}}V,\, \bar{\mathcal{D}}X \rrangle.
\end{equation}
Because of 
\begin{equation}\label{vx}
V X V^{-1} = \omega EV^2 X \quad \mbox{with} \quad
\omega EV^2 \in \bar{\mathcal{D}},
\end{equation}
it follows that $\bar{\mathcal{D}}V$ and $\bar{\mathcal{D}}X$ commute.
Moreover, since $V^2$ and $X^2$ are both elements of
$\bar{\mathcal{D}}$, the right-hand side of equation~(\ref{sbb54}) is
isomorphic to $\zz_2 \times \zz_2$, which is Klein's four-group.

The case of $\sbb/\Delta(27)$ is more involved. We denote the set 
of matrices of $\Delta(27)$ by $\mathcal{D}$. Therefore, we find  
\begin{equation}
\sbb/\Delta(27) = \llangle \mathcal{D}V,\, \mathcal{D}X \rrangle.
\end{equation}
From equation~(\ref{ordX}) we deduce
\begin{equation}\label{V22}
(\mathcal{D}V)^2 = (\mathcal{D}X)^2 = \mathcal{D}V^2, 
\quad
(\mathcal{D}V)^4 = (\mathcal{D}X)^4 = \mathcal{D}.
\end{equation}
Moreover, equation~(\ref{ordX}) tells us that $\mathcal{D}V^2$ belongs to
the center of $\sbb/\Delta(27)$. Application of equation~(\ref{vx})
leads to
\begin{equation}\label{DVX}
(\mathcal{D}V) (\mathcal{D}X) = 
(\mathcal{D}V^2) (\mathcal{D}X) (\mathcal{D}V).
\end{equation}
This suggests a mapping from $\mathcal{D}V$, 
$\mathcal{D}X$, $\mathcal{D}V^2$ to 
$i\sigma_1$, $i\sigma_2$ and $-\bone_2$, respectively, 
with the Pauli matrices $\sigma_k$ ($k=1,2,3$).
Indeed, with formulas~(\ref{V22}) and~(\ref{DVX}) one can show that 
\begin{equation}\label{qqq}
\begin{array}{l@{\,\to\,}rl@{\,\to\,}rl@{\,\to\,}rl@{\,\to\,}r}
\mathcal{D} & \bone_2, & \mathcal{D}V & i\sigma_1, &
\mathcal{D}X & i\sigma_2, & \mathcal{D}V^3X & i\sigma_3, \\
\mathcal{D}V^2 & -\bone_2, & \mathcal{D}V^3 & -i\sigma_1, &
\mathcal{D}V^2X & -i\sigma_2, & \mathcal{D}VX & -i\sigma_3\hphantom{,}
\end{array}
\end{equation}
constitutes an isomorphism between $\sbb/\Delta(27)$ and the group 
\begin{equation}\label{q8}
Q_8 = \{ \pm \bone_2,\, 
\pm i\sigma_1,\, \pm i\sigma_2,\, \pm i\sigma_3 \}.
\end{equation}
For general discussions, model building and references on the groups
$Q_{2n}$ see~\cite{kubo}. 
\begin{table}
\begin{center}
\begin{tabular}{||c||ccccc||}
\hline \hline
$Q_{8}$ & 
$C^q_1$ & $C^q_2$ & $C^q_3$ & $C^q_4$ & $C^q_5$ \\
(\# $C_k$) & (1) & (1) & (2) & (2) & (2)\\
$\mathrm{ord}(C_k)$ & 1 & 2 & 4 & 4 &  4 \\
\hline \hline
$\mathbf{1}^{(0,0)}$
& $1$ & $1$ & $1$ & $1$ & $1$ \\
$\mathbf{1}^{(1,0)}$
& $1$ & $1$ & $-1$ & $1$ & $-1$ \\
$\mathbf{1}^{(0,1)}$
& $1$ & $1$ & $1$ & $-1$ & $-1$ \\
$\mathbf{1}^{(1,1)}$
& $1$ & $1$ & $-1$ & $-1$ & $1$ \\
\hline
$\mathbf{2}$
& $2$ & $-2$ & $0$ & $0$ & $0$ \\
\hline \hline
\end{tabular}
\caption{Character table of $Q_{8}$.}\label{ctsq8}
\end{center}
\end{table}
The group $Q_8$ has five classes:
\begin{equation}
C^q_1 = \{ \bone_2 \},\; C^q_2 = \{ -\bone_2 \},\;
C^q_3 = \{ \pm i\sigma_1 \},\; C^q_4 = \{ \pm i\sigma_2 \},\;
C^q_5 = \{ \pm i\sigma_3 \}.
\end{equation}
There are four one-dimensional irreps
\begin{equation}\label{q81}
\pm \bone_2 \to 1, \quad 
\pm i\sigma_1 \to (-1)^p, \quad
\pm i\sigma_2 \to (-1)^q, \quad
\pm i\sigma_3 \to (-1)^{p+q}
\end{equation}
with $p,q = 0,1$. The remaining irrep of $Q_8$ is the defining
two-dimensional irrep~(\ref{q8}).

\subsection{The irreps of $\sbb$}
The first two factor groups in equation~(\ref{f72x3}) furnish the
one-dimensional irreps of $\sbb$:
\begin{equation}\label{1-dim-72x3}
\one^{(p,q)}: \quad C \to 1, \quad E \to 1, \quad
V \to (-1)^p, \quad X \to (-1)^q
\quad (p,q =0,1).
\end{equation}
These correspond to 
the one-dimensional irreps of the factor group $Q_8$.
The two-dimen\-si\-onal irrep of $Q_8$ 
can be translated according to equation~(\ref{qqq}) 
into an irrep of $\sbb$ by 
\begin{equation}
\two: \quad C \to \bone_2, \quad E \to \bone_2, \quad 
V \to i\sigma_1, \quad X \to i\sigma_2.
\end{equation}

The last factor group is $\Sigma(72)$. Its classes are almost trivially found
from those of $\sbb$:
\begin{equation}
\begin{array}{ccl}
{\widetilde C}'_1 &=& \mathcal{Z}, \\
{\widetilde C}'_2 &=& C'_4 \mathcal{Z}, \\
{\widetilde C}'_3 &=& C'_5 \mathcal{Z} \equiv C'_6 \mathcal{Z} \equiv 
C'_7 \mathcal{Z}, \\
{\widetilde C}'_4 &=& C'_8 \mathcal{Z} \equiv C'_9 \mathcal{Z} \equiv 
C'_{10} \mathcal{Z}, \\
{\widetilde C}'_5 &=& C'_{11} \mathcal{Z} \equiv C'_{12} \mathcal{Z} \equiv 
C'_{13} \mathcal{Z}, \\
{\widetilde C}'_6 &=& C'_{14} \mathcal{Z} \equiv C'_{15} \mathcal{Z} \equiv 
C'_{16} \mathcal{Z}.
\end{array}
\end{equation}
Since the one and two-dimensional irreps of $\sbb$ which we have already
constructed have a trivial center, there remains one irrep of
$\Sigma(72)$. Because of $4 \times 1^2 + 2^2 + 8^2 = 72$, this irrep must have
dimension eight. If we denote---for reasons to become clear soon---the defining
irrep of $\sbb$ by $\three^{(0,0)}$, this irrep $\eight$ must be obtained by
\begin{equation}
\three^{(0,0)} \otimes \left( \three^{(0,0)} \right)^* = 
\one \oplus \eight.
\end{equation}
For $\saa$ the eight-dimensional representation decayed into two
four-dimensional irreps---see appendix~\ref{tensor3}. However, $\sbb$ has one
generator more and the $\eight$ is irreducible, as one can easily show. 
The character of the $\eight$ is thus given by
\begin{equation}
\chi_\eight = \left| \chi_{\three^{(0,0)}} \right|^2 - 1,
\end{equation}
which allows to complete the character table of $\Sigma(72)$.
\begin{table}
\begin{center}
\begin{tabular}{||c||cccccc||}
\hline \hline
$\Sigma(72)$\rule{0mm}{15pt} & ${\widetilde C}'_1$ & ${\widetilde C}'_2$ & 
${\widetilde C}'_3$ & ${\widetilde C}'_4$ & ${\widetilde C}'_5$ & 
${\widetilde C}'_6$ \\
(\# $C_k$) & (1) & (8) & (9) & (18) & (18) & (18) \\
$\mathrm{ord}(C_k)$ & 1 & 3 & 2 & 4 &  4 & 4\\
\hline \hline
$\mathbf{1}^{(1)}$
& $1$ & $1$ & $1$ & $1$ & $1$ & $1$ \\
$\mathbf{1}^{(2)}$
& $1$ & $1$ & $1$ & $-1$ & $1$ & $-1$ \\
$\mathbf{1}^{(3)}$
& $1$ & $1$ & $1$ & $1$ & $-1$ & $-1$ \\
$\mathbf{1}^{(4)}$
& $1$ & $1$ & $1$ & $-1$ & $-1$ & $1$ \\
\hline
$\mathbf{2}$
& $2$ & $2$ & $-2$ & $0$ & $0$ & $0$ \\
\hline
$\mathbf{8}$
& $8$ & $-1$ & $0$ & $0$ & $0$ & $0$ \\
\hline \hline
\end{tabular}
\caption{Character table of $\Sigma(72)$.}\label{cts72}
\end{center}
\end{table}

Now we have exhausted the factor groups. The ten remaining irreps must be
faithful. We readily find eight three-dimensional irreps by multiplying the
defining irrep with the one-dimensional irreps and by complex conjugation: 
\begin{equation}
\three^{(p,q)} \equiv \one^{(p,q)} \otimes \three^{(0,0)}, 
\quad
\left( \three^{(p,q)} \right)^*  \equiv \one^{(p,q)} \otimes 
\left( \three^{(0,0)} \right)^*.
\end{equation}
Explicitly, the generators of $\sbb$ are represented by
\begin{equation}
\begin{array}{ccccc}
\three^{(p,q)}: &
C \to C, & E \to E, & V \to (-1)^p V, & X \to (-1)^q X, \\
\left( \three^{(p,q)} \right)^*: &
C \to C^*, & E \to E, & V \to (-1)^p V^*, & X \to (-1)^q X^*.
\end{array}
\end{equation}
\begin{table}
\begin{small}
\begin{center}
\rotatebox{90}{
\begin{tabular}{||c||ccc|c|ccc|ccc|ccc|ccc||}
\hline \hline
$\sbb$ & 
$C_1'$ & $C_2'$ & $C_3'$ & $C_4'$ & $C_5'$ & $C_6'$ &
$C_7'$ & $C_8'$ & 
$C_9'$ & $C_{10}'$ & $C'_{11}$ & $C'_{12}$ & $C'_{13}$
& $C'_{14}$ & $C'_{15}$ & $C'_{16}$ \\ 
(\# $C_k$) & (1) & (1) & (1) & (24) & (9) & (9) & (9) & (18) &
(18) & (18) & (18) & (18) & (18) & (18) & (18) & (18) \\
$\mathrm{ord}(C_k)$ & 1 & 3 & 3 & 3 & 2 & 6 & 6 & 4 & 12 & 12 & 4 & 12 & 12 &
12 & 12 & 4 \\
\hline \hline
$\mathbf{1}^{(0,0)}$ & $1$ & $1$ & $1$ & $1$ & $1$ 
& $1$ & $1$ & $1$
& $1$ & $1$ & $1$
& $1$ & $1$ & $1$ & $1$ & $1$ \\
$\mathbf{1}^{(1,0)}$ & $1$ & $1$ & $1$ & $1$ & $1$ 
& $1$ & $1$ & $-1$
& $-1$ & $-1$ & $1$ 
& $1$ & $1$ & $-1$ & $-1$ & $-1$ \\
$\mathbf{1}^{(0,1)}$ &
$1$ & $1$ & $1$ & 
$1$ &
$1$ & $1$ & $1$ &
$1$ & $1$ & $1$ &
$-1$ & $-1$ & $-1$ &
$-1$ & $-1$ & $-1$ \\
$\mathbf{1}^{(1,1)}$ &
$1$ & $1$ & $1$ & 
$1$ &
$1$ & $1$ & $1$ &
$-1$ & $-1$ & $-1$ &
$-1$ & $-1$ & $-1$ &
$1$ & $1$ & $1$ \\
\hline
$\mathbf{2}$ &
$2$ & $2$ & $2$ & 
$2$ &
$-2$ & $-2$ & $-2$ &
$0$ & $0$ & $0$ &
$0$ & $0$ & $0$ &
$0$ & $0$ & $0$ \\
\hline
$\mathbf{3}^{(0,0)}$ &
$3$ & $3\omega$ & $3\omega^2$ & 
$0$ &
$-1$ & $-\omega$ & $-\omega^2$ &
$1$ & $\omega$ & $\omega^2$ &
$1$ & $\omega$ & $\omega^2$ &
$\omega$ & $\omega^2$ & $1$ \\
$\mathbf{3}^{(1,0)}$ &
$3$ & $3\omega$ & $3\omega^2$ & 
$0$ &
$-1$ & $-\omega$ & $-\omega^2$ &
$-1$ & $-\omega$ & $-\omega^2$ &
$1$ & $\omega$ & $\omega^2$ &
$-\omega$ & $-\omega^2$ & $-1$ \\
$\mathbf{3}^{(0,1)}$ &
$3$ & $3\omega$ & $3\omega^2$ & 
$0$ &
$-1$ & $-\omega$ & $-\omega^2$ &
$1$ & $\omega$ & $\omega^2$ &
$-1$ & $-\omega$ & $-\omega^2$ &
$-\omega$ & $-\omega^2$ & $-1$ \\
$\mathbf{3}^{(1,1)}$ &
$3$ & $3\omega$ & $3\omega^2$ & 
$0$ &
$-1$ & $-\omega$ & $-\omega^2$ &
$-1$ & $-\omega$ & $-\omega^2$ &
$-1$ & $-\omega$ & $-\omega^2$ &
$\omega$ & $\omega^2$ & $1$ \\
$\left( \mathbf{3}^{(0,0)} \right)^*$ &
$3$ & $3\omega^2$ & $3\omega$ & 
$0$ &
$-1$ & $-\omega^2$ & $-\omega$ &
$1$ & $\omega^2$ & $\omega$ &
$1$ & $\omega^2$ & $\omega$ &
$\omega^2$ & $\omega$ & $1$ \\
$\left( \mathbf{3}^{(1,0)} \right)^*$ &
$3$ & $3\omega^2$ & $3\omega$ & 
$0$ &
$-1$ & $-\omega^2$ & $-\omega$ &
$-1$ & $-\omega^2$ & $-\omega$ &
$1$ & $\omega^2$ & $\omega$ &
$-\omega^2$ & $-\omega$ & $-1$ \\
$\left( \mathbf{3}^{(0,1)} \right)^*$ &
$3$ & $3\omega^2$ & $3\omega$ & 
$0$ &
$-1$ & $-\omega^2$ & $-\omega$ &
$1$ & $\omega^2$ & $\omega$ &
$-1$ & $-\omega^2$ & $-\omega$ &
$-\omega^2$ & $-\omega$ & $-1$ \\
$\left( \mathbf{3}^{(1,1)} \right)^*$ &
$3$ & $3\omega^2$ & $3\omega$ & 
$0$ &
$-1$ & $-\omega^2$ & $-\omega$ &
$-1$ & $-\omega^2$ & $-\omega$ &
$-1$ & $-\omega^2$ & $-\omega$ &
$\omega^2$ & $\omega$ & $1$ \\
\hline
$\mathbf{6}$ &
$6$ & $6\omega$ & $6\omega^2$ & 
$0$ &
$2$ & $2\omega$ & $2\omega^2$ &
$0$ & $0$ & $0$ &
$0$ & $0$ & $0$ &
$0$ & $0$ & $0$ \\
$\mathbf{6^{\ast}}$ &
$6$ & $6\omega^2$ & $6\omega$ & 
$0$ &
$2$ & $2\omega^2$ & $2\omega$ &
$0$ & $0$ & $0$ &
$0$ & $0$ & $0$ &
$0$ & $0$ & $0$ \\
\hline
$\mathbf{8}$ &
$8$ & $8$ & $8$ & 
$-1$ &
$0$ & $0$ & $0$ &
$0$ & $0$ & $0$ &
$0$ & $0$ & $0$ &
$0$ & $0$ & $0$ \\
\hline \hline
\end{tabular}
}
\caption{Character table of $\sbb$.
\label{charactertable72}}
\end{center}
\end{small}
\end{table}

There are two missing irreps with dimensions $d$ and $d'$, which fulfill
$d^2 + {d'}^2 = 72$. The solution of this equation is unique: $d = d' = 6$.
Therefore, the six-dimensional irreps are obtained by the tensor product of
the defining irrep with itself. Again, just as for the $\eight$, the $\six$
and $\six^*$ are irreducible because of the generator $X$ in addition
to those of $\saa$. We define the $\six^*$ by
\begin{equation}\label{3336}
\three^{(0,0)} \otimes \three^{(0,0)} = 
\left( \three^{(0,0)} \right)^* \oplus \six^*
\end{equation}
and the $\six$ by its complex conjugate.
This completes the construction of all irreps of $\sbb$. 
With equation~(\ref{3336}), the
character of the $\six^*$ is then given by
\begin{equation}
\chi_{\six^*} = \left( \chi_{\three^{(0,0)}} \right)^2 -  
\left( \chi_{\three^{(0,0)}} \right)^*.
\end{equation}
With $\chi_\six = \left( \chi_{\six^*} \right)^*$, we can 
complete the character table of $\sbb$---see 
table~\ref{charactertable72}. 

\section{The group $\scc$}
\label{S216}

\subsection{Definition, principal series and conjugacy classes}

The so-called Hessian group $\scc$ is generated by
\begin{equation}
\scc = \langle\langle\, C,\, E,\, V,\, D\, \rangle\rangle.
\end{equation}
Because of $X = D V D^{-1}$, also $X$ is an element of $\scc$.
In order to prove the principal series~(\ref{csS216}), we use the
following relations:
\begin{subequations}
\begin{eqnarray}
D^{-1} C D = C, \; D^{-1} E D = CE &\Rightarrow& \Delta(27) \llhd
\scc, \\
D^{-1} V^2 D = CV^2 &\Rightarrow& \Delta(54) \llhd \scc, \\
D^{-1} V D = \omega^2 CV^3X,\; D^{-1} X D = V &\Rightarrow& 
\sbb \llhd \scc. \label{d3}
\end{eqnarray}
\end{subequations}
With our knowledge about the principal series of $\sbb$ this proves
equation~(\ref{csS216}), the principal series of $\scc$. 
The group $\saa$ is not a member of this
principal series because of the first relation in equation~(\ref{d3});
moreover, this is also clear from the discussion at the end of
section~\ref{subsbb}. 

Further relations for the determination of the classes of $\scc$ are
\begin{equation}
C^{-1} D C = D, \quad E^{-1} D E = \omega C^2 D, \quad
V^{-1} D V = V^3 D V = V^3XD.
\end{equation}
Note that without $X$ the expression $V^3 D V$ cannot be reordered
anymore. The same applies to $D^{-1} V D = CV^3DVD^2 = \omega^2
CV^3X$. This is the reason that the elements of $\scc$ cannot be
written as $\omega^\alpha C^\beta E^\gamma V^\delta D^\epsilon$ with 
$\alpha, \beta, \gamma = 0,1,2$, $\delta = 0,1,2,3$ and $\epsilon =
0,1,2$. (Note that $D^3 = \omega^2 \bone$.) Actually, if this were the
case, the number of elements of this group would be too small:
$3 \times 3 \times 3 \times 4 \times 3 = 324$. However, every element
$g \in \scc$ can be written as
\begin{equation}
g = \omega^p C^q E^r V^s X^t D^u \quad \mbox{with} \quad
p,q,r,u = 0,1,2, \; s = 0,1,2,3,\; t = 0,1,
\end{equation}
which gives the correct number of elements.

In order to find the conjugacy classes of $\scc$ one could use
theorems~\ref{1} and~\ref{2} and the conjugacy classes
of~$\sbb$. However, since $\scc$ has 648 elements, 
we use GAP~\cite{gap} to find the classes and confine ourselves to 
checking the result. The list of classes is displayed in
table~\ref{classes-216x3}. For the classes $C''_k$ of $\scc$ which are subsets
of $\sbb$ we have indicated the classes $C'_l$ of $\sbb$ they consist
of. Furthermore, we have characterized every $C''_k$ by 
one element $g \in C''_k$.
\begin{table}
	\begin{center}
	\begin{tabular}{|r@{\,=\,}l|r|}
	\hline
\multicolumn{2}{|c}{Class} & Number of elements\\
	\hline\hline
\hspace{10mm}
$C_1''$ & $C_1' = C_\bone$ & 1\hspace{15mm} \\ 
$C_2''$ & $\omega C_1''$ & 1\hspace{15mm} \\ 
$C_3''$ & $\omega^2 C_1''$ & 1\hspace{15mm} \\
$C_4''$ & $C_4' = C_C$ & 24\hspace{15mm} \\
$C_5''$ & $C_5' = C_{V^2}$ & 9\hspace{15mm} \\
$C_6''$ & $\omega C_5'$ & 9\hspace{15mm} \\
$C_7''$ & $\omega^2 C_5'$ & 9\hspace{15mm} \\
$C_8''$ & $C_8'\cup C'_{11}\cup C'_{16} = C_V$ & 54\hspace{15mm} \\ 
$C_9''$ & $\omega C_8''$ & 54\hspace{15mm} \\ 
$C_{10}''$ & $\omega^2 C_8''$ & 54\hspace{15mm} \\ 
	\hline
$C_{11}''$ & $C_{ED}$ & 72\hspace{15mm} \\
$C_{12}''$ & $C_{ED^2}$ & 72\hspace{15mm} \\
$C_{13}''$ & $C_D$ & 12\hspace{15mm} \\
$C_{14}''$ & $\omega C_{D}$ & 12\hspace{15mm} \\
$C_{15}''$ & $\omega^2 C_{D}$ & 12\hspace{15mm} \\
$C_{16}''$ & $C_{D^2}$ & 12\hspace{15mm} \\
$C_{17}''$ & $\omega C_{D^2}$ & 12\hspace{15mm} \\
$C_{18}''$ & $\omega^2 C_{D^2}$ & 12\hspace{15mm} \\
$C_{19}''$ & $C_{V^2 D}$ & 36\hspace{15mm} \\
$C_{20}''$ & $\omega C_{V^2 D}$ & 36\hspace{15mm} \\
$C_{21}''$ & $\omega^2 C_{V^2 D}$ & 36\hspace{15mm} \\
$C_{22}''$ & $C_{VD^2}$ & 36\hspace{15mm} \\
$C_{23}''$ & $\omega C_{VD^2}$ & 36\hspace{15mm} \\
$C_{24}''$ & $\omega^2 C_{VD^2}$ & 36\hspace{15mm} \\
	\hline
\multicolumn{2}{|c}{Total number of elements} & 648\hspace{15mm} \\
	\hline
	\end{tabular}
	\end{center}
\caption{The conjugacy classes of $\scc$. \label{classes-216x3}}
\end{table}

One can show, by proceeding as described at the end of
section~\ref{D27}, that $\scc$ has no other proper
normal subgroups than those appearing in its principal series.

\subsection{The factor groups}

In order to characterize the factor groups in the principal
series~(\ref{csS216}), we first have to discuss the groups $T'$ and
$A_4$, as will shortly become clear.

According to GAP, the group $T'$, the double covering group of
$A_4$, has a presentation with two generators $a$, $b$ as
\begin{equation}\label{pres1}
T': \quad a^4 = a^2 b^{-3} = (ab)^3 = e.
\end{equation}
Obviously, the element
\begin{equation}\label{v}
v \equiv a^2 = b^3
\end{equation}
belongs to the center of $T'$.
From the presentation~(\ref{pres1}), 
the group $A_4$ is obtained by the restriction
$v= a^2 = b^3 = e$, \textit{i.e.}
\begin{equation}
A_4: \quad a^2 = b^3 = (ab)^3 = e.
\end{equation}
It is easy to show that Klein's four-group $\zz_2 \times \zz_2$ 
is a normal subgroup of $A_4$ generated by the commuting elements
$a$ and $b a b^{-1}$. Therefore, $A_4$ has the structure 
$(\zz_2 \times \zz_2) \rtimes \zzz$ where $b$ generates the $\zzz$.

Alternatively, one can use the generators $s = a$, $t = ab$ for a
presentation of $T'$. These fulfill~\cite{hagedorn}
\begin{equation}\label{pres2}
T': \quad s^4 = t^3 = (st)^3 = e.
\end{equation}
In equation~(\ref{pres2}) the step from  $T'$ to $A_4$
requires $s^2 = e$.

Now we establish the isomorphisms
\begin{equation}\label{fS216}
\scc/\Delta(27) \cong T', \quad \scc/\Delta(54) \cong A_4, \quad 
\scc/\sbb \cong \zzz.
\end{equation}
With $\mathcal{D}$ denoting again the set of elements of $\Delta(27)$, 
the assignment 
\begin{equation}\label{assign1}
s \to \mathcal{D} V, \quad t \to \mathcal{D} D
\end{equation}
proves the isomorphism $\scc/\Delta(27) \cong T'$, because 
$V^4 = (VD)^3 = \bone$ and $D^3 = \omega^2 \bone \in \mathcal{D}$. 
Consequently, the
second relation of equation~(\ref{fS216}) is proven by the assignment
\begin{equation}\label{assign2}
s \to \bar{\mathcal{D}} V, \quad t \to \bar{\mathcal{D}} D,
\end{equation}
where $\bar{\mathcal{D}}$ is the set of elements of
$\Delta(54)$. Since $V^2 \in \bar{\mathcal{D}}$ and, therefore, 
$(\bar{\mathcal{D}} V)^2 = \bar{\mathcal{D}}$, the
assignment~(\ref{assign2}) establishes the isomorphism between $A_4$
and $\scc/\Delta(54)$. The third relation in equation~(\ref{fS216})
follows trivially from
\begin{equation}\label{assign3}
\scc/\sbb = \llangle \widetilde{\mathcal{D}} D \rrangle,
\end{equation}
where $\widetilde{\mathcal{D}}$ denotes the sets of elements of $\sbb$.

Now, for the purpose of applying the result to $\scc$, we develop the
representation theory of $A_4$ and $T'$. The relationship between
these group is the same as between $SO(3)$ and $SU(2)$. Given a
rotation matrix $R \in SO(3)$ then there are exactly two $SU(2)$
matrices $U$ which differ only in the overall sign such that
\begin{equation}\label{su2-so3}
U \left( \vec \sigma \cdot \vec x \right) U^\dagger = 
\vec \sigma \cdot \left( R \vec x \right)
\end{equation}
for all vectors $\vec x \in \mathbbm{R}^3$. We use the notation
$\vec \sigma \cdot \vec x \equiv \sum_{k=1}^3 \sigma_k x_k$ with the
Pauli matrices $\sigma_k$. 
Vice versa, any $SU(2)$ matrix $U$ induces a rotation on
$\mathbbm{R}^3$ via equation~(\ref{su2-so3}).

According to the definition of $A_4$ in section~\ref{introduction}, 
this group is generated by the $SO(3)$ matrices
\begin{equation}\label{Rab}
R_a = A, \quad R_b = E,
\end{equation}
which fulfill all relations of the presentation of $A_4$ mentioned
above. The corresponding matrices $U_a$ and $U_b$ have to be found
through equation~(\ref{su2-so3}). In contrast to $SU(2)$, the signs of
these matrices are fixed by the additional relations of the
presentation of $T'$. Firstly, $U_a^2$ is in the center, therefore, 
$U_a^2 = -\bone_2 = U_b^3$ and, secondly, $(U_a U_b)^3 = \bone_2$.
The result of the computation is
\begin{equation}\label{Uab}
U_a = i \left( \begin{array}{cc} 0 & 1 \\ 1 & 0 
\end{array} \right), \quad
U_b = \frac{1}{\sqrt{2}} \left( 
\begin{array}{cc} \phi & \phi \\ -\phi^* & \phi^* 
\end{array} \right)
\quad \mbox{with} \quad \phi = e^{i\pi/4}.
\end{equation}
Equations~(\ref{assign1}) and (\ref{assign2}) suggest to use
\begin{equation}
U_t \equiv U_a U_b = \frac{1}{\sqrt{2}} \left( 
\begin{array}{cc} -\phi & \phi \\ -\phi^* & -\phi^* 
\end{array} \right)
\end{equation}
instead of $U_b$.
In order to construct all irreps of $T'$, it is useful to
first compute its classes. With the relation of the presentation we obtain
\begin{equation}
\begin{array}{ccccl}
&& C^t_1 &=& \{ e \}, \\
&& C^t_2 &=& \{ v \}, \\
s & \in & C^t_3 &=& \{ a,\, va,\, bab^2,\, vbab^2,\, b^2ab,\, vb^2ab \}, \\ 
t & \in & C^t_4 &=& \{ ab,\, ba,\, vb,\, aba \}, \\
t^2& \in & C^t_5 &=& \{ ab^2,\, b^2a,\, b^2,\, bab \}, \\
vt & \in & C^t_6 &=& \{ vab,\, vba,\, b,\, vaba \}, \\
vt^2& \in & C^t_7 &=& \{ vab^2,\, vb^2a,\, vb^2,\, vbab \}
\end{array}
\end{equation}
Thus we know that $T'$ has seven irreps. The one-dimensional irreps are
given by
\begin{equation}\label{t1}
\one^{(p)}: \quad s \to 1,\quad t \to \omega^p \quad (p=0,1,2).
\end{equation}
We have already found a two-dimensional irrep given by $U_a$ and $U_b$
(or $U_t$) which we denote by $\two^{(0)}$. 
The other two-dimensional irreps are given by
$\two^{(p)} = \one^{(p)} \otimes \two^{(0)}$.
Therefore, we have
\begin{equation}
\two^{(p)}: 
\quad s \to U_a, \quad t \to \omega^p \, U_t \quad (p=0,1,2).
\end{equation}
There remains a three-dimensional irrep which, according to 
$T'/\zz_2 \cong A_4$, must be the one given by
the matrices of $A_4$:
\begin{equation}\label{t3}
\three^{(a)}: s \to R_a = \mbox{diag}\,(1,-1,-1), \quad
t \to R_a R_b = \left( 
\begin{array}{ccc}
0 & 1 & 0 \\ 0 & 0 & -1 \\ -1 & 0 & 0 
\end{array} \right).
\end{equation}
Knowledge of all classes and irreps allows to write down the
character table of $T'$---see table~\ref{ctsl23}.
Further information on this group can be found, for instance,
in~\cite{hagedorn,kephart}. 
\begin{table}
\begin{center}
\begin{small}
\begin{tabular}{||l||ccccccc||}
\hline \hline
$T'$  & $C^t_1$ &  $C^t_2$ & $C^t_3$ & $C^t_4$ & $C^t_5$ & $C^t_6$ & $C^t_7$ \\
(\# $C_k$) & (1) & (1) & (6) & (4) & (4) & (4) &
(4)\\
$\mathrm{ord}(C_k)$ & 1 & 2 & 4 & 3 & 3 & 6 & 6 \\
\hline \hline
$\mathbf{1}^{(0)}$ & 1 & 1 & 1 & 1 & 1 & 1 & 1\\
$\mathbf{1}^{(1)}$ & 1 & 1 & 1 & $\omega$ & $\omega^2$ & $\omega$ & $\omega^2$\\
$\mathbf{1}^{(2)}$ & 1 & 1 & 1 & $\omega^2$ & $\omega$ & $\omega^2$ & $\omega$\\
\hline
$\mathbf{2}^{(0)}$ & 2 & $-2$ & $0$ & $-1$ & $-1$ & $1$ & $1$\\
$\mathbf{2}^{(1)}$ & 2 & $-2$ & $0$ & $-\omega$ & $-\omega^2$ &
$\omega$ & $\omega^2$\\
$\mathbf{2}^{(2)}$ & 2 & $-2$ & $0$ & $-\omega^2$ & $-\omega$ &
$\omega^2$ & $\omega$\\
\hline
$\mathbf{3}^{(a)}$ & 3 & 3 & $-1$ & 0 & 0 & 0 & 0\\
\hline \hline
\end{tabular}
\caption{Character table of $T'$. \label{ctsl23}}
\end{small}
\end{center}
\end{table}

Having complete information on $T'$, the discussion of $A_4$ is trivial.
The classes of $A_4$ are obtained from those of $T'$ by setting $v=e$
and deleting the elements and classes which occur twice:
\begin{equation}
\begin{array}{ccccl}
&& C^a_1 &=& \{ e \}, \\
s & \in & C^a_2 &=& \{ a,\, bab^2,\, b^2ab \}, \\ 
t & \in & C^a_3 &=& \{ ab,\, ba, b,\, aba \}, \\
t^2& \in & C^a_4 &=& \{ ab^2,\, b^2a,\, b^2,\, bab \}
\end{array}
\end{equation}
The irreps are given by equations~(\ref{t1}) and~(\ref{t3}).
For the sake of completeness we present the character table of $A_4$ in
table~\ref{cta4}. 
\begin{table}
\begin{small}
\begin{center}
\begin{tabular}{||l||cccc||}
\hline \hline
$A_4$ & $C^a_1$ &  $C^a_2$ & $C^a_3$ & $C^a_4$ \\
(\# $C_k$) & (1) & (3) & (4) & (4) \\
$\mathrm{ord}(C_k)$ & 1 & 2 & 3 & 3 \\
\hline \hline
$\mathbf{1}^{(0)}$ & 1 & 1 & 1 & 1\\
$\mathbf{1}^{(1)}$ & 1 & 1 & $\omega$ & $\omega^2$ \\
$\mathbf{1}^{(2)}$ & 1 & 1 & $\omega^2$ & $\omega$ \\
\hline
$\mathbf{3}^{(a)}$ & 3 & $-1$ & 0 & 0 \\
\hline \hline
\end{tabular}
\caption{Character table of $A_4$.}\label{cta4}
\end{center}
\end{small}
\end{table}
An interesting feature is that 
$\one^{(p)} \otimes \three^{(a)} \cong \three^{(a)}$.
This follows from the character table~\ref{cta4} due to 
$\chi_{\one^{(p)}} \cdot \chi_{\three^{(a)}} = \chi_{\three^{(a)}}$.

\subsection{Irreps of $\scc$}
\label{irreps scc}

The excursion into $T'$ and $A_4$ nets us seven irreps of~$\scc$, by
utilizing equations~(\ref{assign1}), (\ref{assign2}) and (\ref{assign3}). 
The one-dimensional irreps are given by 
\begin{equation}\label{11}
\one^{(p)}: 
\quad C \to 1 \quad E \to 1, \quad V \to 1, \quad D \to \omega^p
\quad (p=0,1,2).
\end{equation}
The three-dimensional irrep of $A_4$ (or $T'$)~\cite{ludl} leads to 
\begin{equation}\label{33}
\three^{(a)}: 
\quad C \to \bone \quad E \to \bone, \quad V \to R_a, \quad 
D \to R_a R_b.
\end{equation}
Finally, with the two-dimensional irreps of $T'$ we find
\begin{equation}\label{22}
\two^{(p)}: 
\quad C \to \bone_2 \quad E \to \bone_2, \quad V \to U_a, \quad 
D \to \omega^p U_t \quad (p=0,1,2).
\end{equation}

\begin{table}
\begin{small}
\begin{center}
\begin{tabular}{||l||cccccccccc||}
\hline \hline
$\Sigma(216)$ \rule{0mm}{15pt} & 
${\widetilde C}''_1$ & ${\widetilde C}''_2$  &
${\widetilde C}''_3$  & ${\widetilde C}''_4$  & ${\widetilde C}''_5$ &
${\widetilde C}''_6$  & ${\widetilde C}''_7$ & ${\widetilde C}''_8$  &
${\widetilde C}''_9$  & ${\widetilde C}''_{10}$ \\ 
(\# $C_k$) & (1) & (8) & (9) & (54) & (24) & (24) &
(12) & (12) & (36) & (36)\\
$\mathrm{ord}(C_k)$ & 1 & 3 & 2 & 4 & 3 & 3 & 3 & 3 & 6 & 6 \\
\hline \hline
$\mathbf{1}^{(0)}$ & 1 & 1 & 1 & 1 & 1 & 1 & 1 & 1 & 1 & 1\\
$\mathbf{1}^{(1)}$ & 1 & 1 & 1 & 1 & $\omega$ & $\omega^2$ & $\omega$ &
$\omega^2$ & $\omega$ &  $\omega^2$ \\
$\mathbf{1}^{(2)}$ & 1 & 1 & 1 & 1 & $\omega^2$ & $\omega$ & $\omega^2$ &
$\omega$ & $\omega^2$ & $\omega$ \\
\hline
$\mathbf{2}^{(0)}$ & 2 & 2 & $-2$ & $0$ & $-1$ & $-1$ & $-1$ & $-1$ & $1$ &
$1$ \\ 
$\mathbf{2}^{(1)}$ & 2 & 2 & $-2$ & $0$ & $-\omega$ & $-\omega^2$ & $-\omega$
& $-\omega^2$ & $\omega$ &   $\omega^2$ \\ 
$\mathbf{2}^{(2)}$ & 2 & 2 & $-2$ & $0$ & $-\omega^2$ & $-\omega$ &
$-\omega^2$ & $-\omega$ & $\omega^2$ & $\omega$ \\ 
\hline
$\mathbf{3}^{(a)}$ & 3 & 3 & 3 & $-1$ & 0 & 0 & 0 & 0 & 0 & 0\\
\hline
$\mathbf{8}^{(0)}$ & 8 & $-1$ & 0 & 0 & $-1$ & $-1$ & 2 & 2 & 0 & 0\\
$\mathbf{8}^{(1)}$ & 8 & $-1$ & 0 & 0 & $-\omega$ & $-\omega^2$ &
$2\omega$ & $2\omega^2$ & 0 & 0 \\
$\mathbf{8}^{(2)}$ & 8 & $-1$ & 0 & 0 & $-\omega^2$ & $-\omega$ & $2\omega^2$
& $2\omega$ & 0 & 0 \\ 
\hline \hline
\end{tabular}
\caption{Character table of $\Sigma(216)$.}\label{cts216}
\end{center}
\end{small}
\end{table}
Until now we have exploited the first two factor groups of
equation~(\ref{fS216}). It remains to discuss 
$\Sigma(216) = \scc/\zzz$. 
Examining table~\ref{classes-216x3} we find that the 24 classes of
$\scc$ collapse into ten classes of $\Sigma(216)$:
\begin{equation}
\begin{array}{ccl}
{\widetilde C}''_1 &=& \mathcal{Z}, \\
{\widetilde C}''_2 &=& C''_4 \mathcal{Z}, \\
{\widetilde C}''_3 &=& C''_5 \mathcal{Z} \equiv C''_6 \mathcal{Z} \equiv 
C''_7 \mathcal{Z}, \\
{\widetilde C}''_4 &=& C''_8 \mathcal{Z} \equiv C''_9 \mathcal{Z} \equiv 
C''_{10} \mathcal{Z}, \\
{\widetilde C}''_5 &=& C''_{11} \mathcal{Z}, \\
{\widetilde C}''_6 &=& C''_{12} \mathcal{Z}, \\
{\widetilde C}''_7 &=& C''_{13} \mathcal{Z} \equiv C''_{14} \mathcal{Z} \equiv 
C''_{15} \mathcal{Z}, \\
{\widetilde C}''_8 &=& C''_{16} \mathcal{Z} \equiv C''_{17}
\mathcal{Z} \equiv C''_{18} \mathcal{Z}, \\
{\widetilde C}''_9 &=& C''_{19} \mathcal{Z} \equiv C''_{20}
\mathcal{Z} \equiv C''_{21} \mathcal{Z}, \\
{\widetilde C}''_{10} &=& C''_{22} \mathcal{Z} \equiv C''_{23}
\mathcal{Z} \equiv C''_{24} \mathcal{Z}.
\end{array}
\end{equation}
Since the irreps in equations~(\ref{11}), (\ref{33}) and (\ref{22}) map
the center $\zzz$ of $\scc$ onto the unit matrix, these irreps provide
seven irreps of $\Sigma(216)$. There are three remaining ones.
We denote the defining irrep of $\scc$ by $\three^{(0)}$.
Since in the tensor product
\begin{equation}
\three^{(0)} \otimes \left( \three^{(0)} \right)^* = 
\one^{(0)} \oplus \eight^{(0)}
\end{equation}
the center is trivially represented, the $\eight^{(0)}$ is an irrep of
$\Sigma(216)$. Actually there are three eight-dimensional irreps:
\begin{equation}
\eight^{(p)} = \one^{(p)} \otimes \eight^{(0)} \quad (p=0,1,2).
\end{equation}
Since $3 \times 1^2 + 3 \times 2^2 + 3^2 + 3 \times 8^2 = 216$, we have
found all irreps of $\Sigma(216)$---see also~\cite{fairbairn}. 
Its character table is presented in table~\ref{cts216}.

\renewcommand{\thetable}{\arabic{table}a}
\begin{table}
\begin{small}
\begin{center}
\rotatebox{90}{
\begin{tabular}{||c||ccc|c|ccc|ccc|c|c||}
\hline \hline
$\scc$ & 
$C''_1$ & $\omega C''_2$ & $C''_3$ & $C''_4$ & $C''_5$ & $C''_6$ &
$C''_7$ & $C''_8$ & $C''_9$ & $C''_{10}$ & $C''_{11}$ & $C''_{12}$ \\
(\# $C_k$) & (1) & (1) & (1) & (24) & (9) & (9) & (9) & (54) &
(54) & (54) & (72) & (72)\\
$\mathrm{ord}(C_k)$ & 1 & 3 & 3 & 3 & 2 & 6 & 6 & 4 & 12 & 12 & 3 & 3 \\
\hline \hline
$\mathbf{1}^{(0)}$ & 1 & 1 & 1 & 1 & 1 & 1 & 1 & 1 & 1 & 1 & 1 & 1\\
$\mathbf{1}^{(1)}$ & 1 & 1 & 1 & 1 & 1 & 1 & 1 & 1 & 1 & 1 & 
$\omega$ & $\omega^2$ \\
$\mathbf{1}^{(2)}$ & 1 & 1 & 1 & 1 & 1 & 1 & 1 & 1 & 1 & 1 &
$\omega^2$ & $\omega$ \\ 
\hline
$\mathbf{2}^{(0)}$ & 2 & 2 & 2 & 2 & $-2$ & $-2$ & $-2$ &   0 &    0 &
0 & $-1$ & $-1$ \\
$\mathbf{2}^{(1)}$ & 2 & 2 & 2 & 2 & $-2$ & $-2$ & $-2$ &   0 &    0 &
0 &    $-\omega$ &   $-\omega^2$ \\ 
$\mathbf{2}^{(2)}$ & 2 & 2 & 2 & 2 & $-2$ & $-2$ & $-2$ &   0 &    0 &
0 &   $-\omega^2$ &    $-\omega$ \\ 
\hline
$\mathbf{3}^{(a)}$ & 3 & 3 & 3 & 3 & 3 & 3 & 3 & $-1$ & $-1$ & $-1$ &
0 & 0 \\
$\mathbf{3}^{(0)}$ & 3 &   $3\omega$ &  $3\omega^2$ & 0 & $-1$ &
$-\omega$ &  $-\omega^2$ & 1 &   $\omega$ &  $\omega^2$ & 0 &  0 \\
$\mathbf{3}^{(1)}$ & 3 &   $3\omega$ &  $3\omega^2$ &   0 & $-1$ &
$-\omega$ &  $-\omega^2$ & 1 &   $\omega$ &  $\omega^2$ &    0 &    0 \\
$\mathbf{3}^{(2)}$ & 3 &   $3\omega$ &  $3\omega^2$ &   0 & $-1$ &
$-\omega$ &  $-\omega^2$ & 1 &   $\omega$ &  $\omega^2$ &    0 &    0 \\
$\left( \mathbf{3}^{(0)} \right)^*$ & 3 &  $3\omega^2$ &   $3\omega$ &
0 & $-1$ &  $-\omega^2$ &   $-\omega$ & 1 &  $\omega^2$ &   $\omega$ &
0 & 0 \\ 
$\left( \mathbf{3}^{(1)} \right)^*$ & 3 &  $3\omega^2$ &   $3\omega$ &
0 & $-1$ &  $-\omega^2$ &   $-\omega$ & 1 &  $\omega^2$ &   $\omega$ &
0 & 0 \\
$\left( \mathbf{3}^{(2)} \right)^*$ & 3 &  $3\omega^2$ &   $3\omega$ &
0 & $-1$ &  $-\omega^2$ &   $-\omega$ & 1 &  $\omega^2$ &   $\omega$ &
0 & 0 \\
\hline
$\mathbf{6}^{(0)}$ & 6 & $6\omega$ &   $6\omega^2$ &   0 & 2 &
$2\omega$ &  $2\omega^2$ & 0 & 0 & 0 & 0 & 0 \\
$\mathbf{6}^{(1)}$ & 6 &    $6\omega$ &   $6\omega^2$ &   0 & 2 &
$2\omega$ &  $2\omega^2$ &   0 &    0 &    0 &    0 & 0 \\
$\mathbf{6}^{(2)}$ & 6 &    $6\omega$ &   $6\omega^2$ &   0 & 2 &
$2\omega$ &  $2\omega^2$ &   0 &    0 &    0 &    0 & 0 \\
$\left( \mathbf{6}^{(0)} \right)^*$ & 6 &   $6\omega^2$ &    $6\omega$
& 0 & 2 &  $2\omega^2$ & $2\omega$ & 0 & 0 & 0 & 0 & 0 \\ 
$\left( \mathbf{6}^{(1)} \right)^*$ & 6 &   $6\omega^2$ & $6\omega$
& 0 & 2 &  $2\omega^2$ & $2\omega$ & 0 & 0 & 0 & 0 & 0 \\
$\left( \mathbf{6}^{(2)} \right)^*$ & 6 & $6\omega^2$ & $6\omega$ & 0 & 2 &
$2\omega^2$ & $2\omega$ & 0 & 0 & 0 & 0 & 0 \\
\hline
$\mathbf{8}^{(0)}$ & 8 & 8 & 8 & $-1$ & 0 & 0 & 0 & 0 & 0 &
0 & $-1$ & $-1$ \\
$\mathbf{8}^{(1)}$ & 8 & 8 & 8 & $-1$ & 0 & 0 & 0 & 0 & 0 & 0 & 
$-\omega$ & $-\omega^2$ \\
$\mathbf{8}^{(2)}$ & 8 & 8 & 8 & $-1$ & 0 & 0 & 0 & 0 & 0 & 0 & 
$-\omega^2$ & $-\omega$ \\
\hline
$\mathbf{9}$ & 9 &    $9\omega$ &   $9\omega^2$ &   0 & $-3$ &
$-3\omega$ &  $-3\omega^2$ & $-1$ &    $-\omega$ &   $-\omega^2$ &
0 & 0 \\
$\mathbf{9}^{\ast}$ & 9 &   $9\omega^2$ &    $9\omega$ &   0 & $-3$ &
$-3\omega^2$ & $-3\omega$ & $-1$ & $-\omega^2$ & $-\omega$ & 0 & 0
\\
\hline \hline
\end{tabular}
}
\caption{Character table of $\scc$, part 1.
\label{charactertable216a}}
\end{center}
\end{small}
\end{table}
\addtocounter{table}{-1}
\renewcommand{\thetable}{\arabic{table}b}
\begin{table}
\begin{small}
\begin{center}
\rotatebox{90}{
\begin{tabular}{||c||ccc|ccc|ccc|ccc||}
\hline \hline
$\scc$ & 
$C''_{13}$ & $C''_{14}$ & $C''_{15}$ & $C''_{16}$ & $C''_{17}$ &
$C''_{18}$ & $C''_{19}$ & $C''_{20}$ & $C''_{21}$ & $C''_{22}$ &
$C''_{23}$ & $C''_{24}$ \\ 
(\# $C_k$) & (12) & (12) & (12) & (12) & (12) & (12) & (36) & (36) &
(36) & (36) & (36) & (36) \\
$\mathrm{ord}(C_k)$ & 9 & 9 & 9 & 9 & 9 & 9 & 18 & 18 & 18 & 18 & 18 & 18 \\
\hline \hline
$\mathbf{1}^{(0)}$ & 1 & 1 & 1 & 1 & 1 & 1 & 1 & 1 & 1 & 1 & 1 & 1 \\
$\mathbf{1}^{(1)}$ &  $\omega$ & $\omega$ & $\omega$ &
$\omega^2$ &  $\omega^2$ &  $\omega^2$ & $\omega$ &   $\omega$ &
$\omega$ &  $\omega^2$ &  $\omega^2$ &  $\omega^2$ \\
$\mathbf{1}^{(2)}$ &  $\omega^2$ &  $\omega^2$ &  $\omega^2$ &
$\omega$ &   $\omega$ &   $\omega$ &  $\omega^2$ &  $\omega^2$ &
$\omega^2$ &   $\omega$ &   $\omega$ &   $\omega$ \\
\hline
$\mathbf{2}^{(0)}$ & $-1$ & $-1$ & $-1$ & $-1$ & $-1$ & $-1$ & 1 & 1 &
1 & 1 & 1 & 1 \\
$\mathbf{2}^{(1)}$ &    $-\omega$ &    $-\omega$ &    $-\omega$ &
$-\omega^2$ &   $-\omega^2$ &   $-\omega^2$ &   $\omega$ &   $\omega$
& $\omega$ &  $\omega^2$ &  $\omega^2$ &  $\omega^2$ \\
$\mathbf{2}^{(2)}$ & $-\omega^2$ & $-\omega^2$ & $-\omega^2$ &
$-\omega$ &    $-\omega$ &    $-\omega$ &  $\omega^2$ &  $\omega^2$ &
$\omega^2$ &   $\omega$ &   $\omega$ &   $\omega$ \\
\hline
$\mathbf{3}^{(a)}$ & 0 & 0 & 0 & 0 & 0 & 0 & 0 & 0 & 0 & 0 & 0 & 0 \\
$\mathbf{3}^{(0)}$ &   $\omega^2\sigma^{\ast}$ &   $\sigma^{\ast}$ &
$\omega\sigma^{\ast}$ &  $\sigma$ &    $\omega\sigma$ &
$\omega^2\sigma$ &   $-\omega\rho^{\ast}$ &   $-\omega^2\rho^{\ast}$ &
$-\rho^{\ast}$ &    $-\omega^2\rho$ &    $-\rho$ &    $-\omega\rho$ \\
$\mathbf{3}^{(1)}$ &   $\sigma^{\ast}$ &   $\omega\sigma^{\ast}$ &
$\omega^2\sigma^{\ast}$ &    $\omega^2\sigma$ &    $\sigma$ &
$\omega\sigma$ &   $-\omega^2\rho^{\ast}$ &   $-\rho^{\ast}$ &
$-\omega\rho^{\ast}$ &    $-\omega\rho$ &    $-\omega^2\rho$ &
$-\rho$ \\
$\mathbf{3}^{(2)}$ &   $\omega\sigma^{\ast}$ &
$\omega^2\sigma^{\ast}$ &   $\sigma^{\ast}$ &    $\omega\sigma$ &
$\omega^2\sigma$ &    $\sigma$ &   $-\rho^{\ast}$ &
$-\omega\rho^{\ast}$ &   $-\omega^2\rho^{\ast}$ &    $-\rho$ &
$-\omega\rho$ & $-\omega^2\rho$ \\
$\left( \mathbf{3}^{(0)} \right)^*$ & $\omega\sigma$ & $\sigma$ &
$\omega^2\sigma$ &   $\sigma^{\ast}$ &   $\omega^2\sigma^{\ast}$ &
$\omega\sigma^{\ast}$ &    $-\omega^2\rho$ &    $-\omega\rho$ &
$-\rho$ &   $-\omega\rho^{\ast}$ &   $-\rho^{\ast}$ &
$-\omega^2\rho^{\ast}$ \\
$\left( \mathbf{3}^{(1)} \right)^*$ & $\sigma$ &    $\omega^2\sigma$ &
$\omega\sigma$ &   $\omega\sigma^{\ast}$ &   $\sigma^{\ast}$ &
$\omega^2\sigma^{\ast}$ &    $-\omega\rho$ &    $-\rho$ &
$-\omega^2\rho$ &   $-\omega^2\rho^{\ast}$ &   $-\omega\rho^{\ast}$ &
$-\rho^{\ast}$ \\
$\left( \mathbf{3}^{(2)} \right)^*$ &    $\omega^2\sigma$ &
$\omega\sigma$ &    $\sigma$ &   $\omega^2\sigma^{\ast}$ &
$\omega\sigma^{\ast}$ &   $\sigma^{\ast}$ &    $-\rho$ &
$-\omega^2\rho$ &    $-\omega\rho$ &   $-\rho^{\ast}$ &
$-\omega^2\rho^{\ast}$ &   $-\omega\rho^{\ast}$ \\
\hline
$\mathbf{6}^{(0)}$ &  $-\omega\sigma^{\ast}$ &
$-\omega^2\sigma^{\ast}$ &  $-\sigma^{\ast}$ &   $-\omega\sigma$ &
$-\omega^2\sigma$ &   $-\sigma$ &   $-\rho^{\ast}$ &
$-\omega\rho^{\ast}$ &   $-\omega^2\rho^{\ast}$ &    $-\rho$ &
$-\omega\rho$ &    $-\omega^2\rho$ \\
$\mathbf{6}^{(1)}$ &  $-\omega^2\sigma^{\ast}$ &  $-\sigma^{\ast}$ &
$-\omega\sigma^{\ast}$ &   $-\sigma$ &   $-\omega\sigma$ &
$-\omega^2\sigma$ &   $-\omega\rho^{\ast}$ &   $-\omega^2\rho^{\ast}$
&   $-\rho^{\ast}$ &    $-\omega^2\rho$ &    $-\rho$ &
$-\omega\rho$ \\
$\mathbf{6}^{(2)}$ &  $-\sigma^{\ast}$ &  $-\omega\sigma^{\ast}$ &
$-\omega^2\sigma^{\ast}$ &   $-\omega^2\sigma$ &   $-\sigma$ &
$-\omega\sigma$ &   $-\omega^2\rho^{\ast}$ &   $-\rho^{\ast}$ &
$-\omega\rho^{\ast}$ &    $-\omega\rho$ &    $-\omega^2\rho$ & $-\rho$
\\
$\left( \mathbf{6}^{(0)} \right)^*$ & $-\omega^2\sigma$ &
$-\omega\sigma$ &   $-\sigma$ &  $-\omega^2\sigma^{\ast}$ &
$-\omega\sigma^{\ast}$ &  $-\sigma^{\ast}$ &    $-\rho$ &
$-\omega^2\rho$ &    $-\omega\rho$ &   $-\rho^{\ast}$ &
$-\omega^2\rho^{\ast}$ &   $-\omega\rho^{\ast}$ \\
$\left( \mathbf{6}^{(1)} \right)^*$ &   $-\omega\sigma$ & $-\sigma$ &
$-\omega^2\sigma$ &  $-\sigma^{\ast}$ &  $-\omega^2\sigma^{\ast}$ &
$-\omega\sigma^{\ast}$ &    $-\omega^2\rho$ &    $-\omega\rho$ &
$-\rho$ &   $-\omega\rho^{\ast}$ &   $-\rho^{\ast}$ &
$-\omega^2\rho^{\ast}$ \\
$\left( \mathbf{6}^{(2)} \right)^*$ &   $-\sigma$ &
$-\omega^2\sigma$ &   $-\omega\sigma$ &  $-\omega\sigma^{\ast}$ &
$-\sigma^{\ast}$ &  $-\omega^2\sigma^{\ast}$ &    $-\omega\rho$ &
$-\rho$ &    $-\omega^2\rho$ &   $-\omega^2\rho^{\ast}$ &
$-\omega\rho^{\ast}$ & $-\rho^{\ast}$ \\
\hline
$\mathbf{8}^{(0)}$ & 2 & 2 & 2 & 2 & 2 & 2 & 0 & 0 & 0 & 0 & 0 & 0 \\
$\mathbf{8}^{(1)}$ &    $2\omega$ &    $2\omega$ &    $2\omega$ &
$2\omega^2$ &   $2\omega^2$ &   $2\omega^2$ &    0 &    0 &    0 &
0 &    0 &    0 \\
$\mathbf{8}^{(2)}$ &   $2\omega^2$ &   $2\omega^2$ &   $2\omega^2$ &
$2\omega$ &    $2\omega$ &    $2\omega$ &    0 &    0 &    0 &    0 &
0 &    0 \\
\hline
$\mathbf{9}$ & 0 & 0 & 0 & 0 & 0 & 0 & 0 & 0 & 0 & 0 & 0 & 0 \\
$\mathbf{9}^{\ast}$ & 0 & 0 & 0 & 0 & 0 & 0 & 0 & 0 & 0 & 0 & 0 & 0 \\
\hline \hline
\end{tabular}
}
\caption{Character table of $\scc$, part 2, with  
  $\rho \equiv e^{2\pi i/9}$, $\sigma \equiv \rho(1+2\omega$). 
\label{charactertable216b}}
\end{center}
\end{small}
\end{table}
We still have to find 14 irreps of $\scc$ which are not irreps of
$\Sigma(216)$. Multiplying the defining irrep by the one-dimensional
irreps and taking complex conjugates we obtain six three-dimensional 
irreps:
\begin{equation}
\three^{(p)} = \one^{(p)} \otimes \three^{(0)},
\quad
\left( \three^{(p)} \right)^* = 
\left( \one^{(p)} \otimes \three^{(0)} \right)^*
\quad (p=0,1,2).
\end{equation}
We can also construct six six-dimensional irreps by
\begin{equation}
\three^{(0)} \otimes \three^{(0)} = 
\left( \three^{(0)} \right)^* \oplus \left( \six^{(0)} \right)^*
\end{equation}
and
\begin{equation}
\six^{(p)} = \one^{(p)} \otimes \six^{(0)},
\quad
\left( \six^{(p)} \right)^* = 
\left( \one^{(p)} \otimes \six^{(0)} \right)^*
\quad (p=0,1,2).
\end{equation}
Finally, according to formula~(\ref{dim-ord}), 
there are two irreps left with dimensions $d$, $d'$ fulfilling 
$d^2 + {d'}^2 = 162$. This equation has the unique solution 
$d = d' = 9$. An explicit construction of these nine-dimensional irreps
is given by
\begin{equation}
\nine = \three^{(0)} \otimes \three^{(a)} 
\quad \mbox{and} \quad
\nine^* = \left( \three^{(0)} \right)^* \otimes \three^{(a)}.
\end{equation}
The proof of the irreducibility of the irrep $\nine$ is presented in
appendix~\ref{nine-216x3}. Thus we have completed the task of
constructing all irreps of $\scc$. Its character table is divided into
two parts, tables~\ref{charactertable216a} 
and~\ref{charactertable216b}, because it does not fit onto one page.
The characters of the six, eight and nine-dimensional irreps are
computed via
\begin{equation}
\chi_{\six^{(p)}} = \chi_{\one^{(p)}} \cdot 
\left[ \left( \chi_{\three^{(0)}}^* \right)^2 - \chi_{\three^{(0)}} \right],
\quad
\chi_{\eight^{(p)}} = \chi_{\one^{(p)}} \cdot 
\left( \left| \chi_{\three^{(0)}} \right|^2 - 1 \right),
\quad
\chi_\nine = \chi_{\three^{(0)}} \cdot \chi_{\three^{(a)}}. 
\end{equation}

\section{Conclusions}
\label{concl}

In this paper we have performed a thorough discussion of the
``exceptional'' finite subgroups $\saa$, $\sbb$ and $\scc$ of $SU(3)$
by means of the concept of principal series. These are maximal chains of
ascending normal subgroups such that each member is a normal subgroup
of all other groups higher up in the chain. Through their principal series
the three groups under discussion have relationships which are useful
for understanding their structures. For instance, all three principal
series contain the sequence $\Delta(27) \llhd \Delta(54)$. 
Using the principal series as a tool we have computed the conjugacy
classes, irreps and character tables. 

For finding the irreps the most useful property of a principal
series~(\ref{principal}) of a group $G$ is that irreps of the factor
groups $G/G_k$ are also irreps of $G$ itself.
Most of the time the
factor groups are relatively uncomplex and it is easy to find their
irreps. Apart from very small abelian factor groups like 
$\zz_2$, $\zzz$ and $\zz_2 \times \zz_2$, among the factor groups 
occurring in the paper there are three interesting 
groups, two of which are widely used in model building, namely
$\sbb/\Delta(27) \cong Q_8$, $\scc/\Delta(27) \cong T'$ and 
$\scc/\Delta(54) \cong A_4$.

Since we have provided the character tables, in principle one can
reduce any tensor product of irreps of the exceptional groups
discussed in the paper. We have explicitly performed the reduction
into irreps for tensor products of three-dimensional
irreps~\cite{ludl}; this could be useful for model building.
Particularly noteworthy are the very unusual Clebsch--Gordan
coefficients occurring in $\three^{(0)} \otimes \three^{(0)}$ of
$\saa$---see~\cite{ludl} and appendix~\ref{tensor3} of the
present paper, and the occurrence of nine-dimensional irreps in the
case of $\scc$.

\paragraph{Acknowledgment:} We thank L.~Lavoura for collaboration in the early
stage of this paper.

\appendix

\section{Semidirect products of groups}
\setcounter{equation}{0}
\renewcommand{\theequation}{A.\arabic{equation}}
\label{semidirect}

Semidirect products are ubiquitous in the theory of finite
groups. Therefore, we present in this appendix the definition and
their most important properties.

Suppose we have two groups $G$ and $H$ and a homomorphism 
$\phi:\, G \rightarrow \mathrm{Aut}(H)$ where $\mathrm{Aut}(H)$ is the
group of automorphisms of $H$, 
\textit{i.e.} the group of isomorphisms $H \rightarrow H$. 
Let us dwell a bit on $\phi$ before we go on to the definition of the
semidirect product $H \rtimes G$. 
We denote the unit elements of $H$, $G$ and $\mathrm{Aut}(H)$
by $e$, $e'$ and $\mbox{id}$, respectively. 
Since $\phi$ is a homomorphism, the
relations $\phi(g_1g_2) = \phi(g_1) \phi(g_2)$ and 
$\phi(e') = \mbox{id}$ hold. Moreover, 
$\phi(g)(h_1h_2) = (\phi(g)h_1) (\phi(g)h_2)$ since $\phi(g)$ is an
automorphism on $H$. 
\begin{define}\label{Def-SDP}
A semidirect product $H \rtimes_\phi G$ of two groups $H$ and $G$ is
defined as the set $H \times G$ with the multiplication law
\begin{equation}\label{sd-mult}
(h_1,g_1)(h_2,g_2) = (h_1\, \phi(g_1)h_2, g_1g_2),
\end{equation}
where $\phi$ is a homomorphism $\phi:\, G \rightarrow \mathrm{Aut}(H)$.
\end{define}
\noindent
Note that the definition depends on the homomorphism $\phi$; via
$\phi$ the group $G$ acts on $H$. For simplicity of notation we will
drop the index $\phi$ at the symbol $\rtimes$ in the following.

The relevant property of $H \rtimes G$ for group theory is the following.
\begin{theorem}\label{P1-SDP}
The semidirect product $H \rtimes G$ laid down in the
definition~\ref{Def-SDP} is a group. 
\end{theorem}
\noindent
To establish the group property we note that $(e,e')$ is the unit
element of $H \rtimes G$ and that
\begin{equation}
(h,g)^{-1} = ( \phi(g^{-1})h^{-1}, g^{-1}).
\end{equation}
Only the verification of associativity of the multiplication law is a
bit lengthier and we leave this as an exercise to the reader.

The multiplication law~(\ref{sd-mult}) is rather abstract. However,
the following discussion attempts to make it more transparent. 
First we note that the relations
\begin{equation}\label{HG1}
(h_1,e') (h_2,e') = (h_1h_2,e') 
\quad \mbox{and} \quad
(e,g_1) (e,g_2) = (e,g_1g_2)
\end{equation}
hold. This means that both $H \times \{e'\}$ and 
$\{e\} \times G$ are subgroups of $H \rtimes G$.
Moreover, due to
\begin{equation}
(h,g) (h',e') (h,g)^{-1} = (h(\phi(g)h')h^{-1},e'),
\end{equation}
$H$ is a normal subgroup.
In addition, any pair $(h,g)$ can uniquely be decomposed into
\begin{equation}\label{HG2}
(h,g) = (h,e') (e,g) = (e,g) ( \bar h,e') 
\quad \mbox{with} \quad \bar h = \phi(g^{-1})h.
\end{equation}
This allows to write the multiplication law~(\ref{sd-mult}) as
\begin{equation}\label{mult1}
(h_1,g_1)(h_2,g_2) = (h_1,e') (e,g_1) (h_2,e')(e,g_1)^{-1} 
(e,g_1) (e,g_2) 
\end{equation}
with 
$(h_1,e') (e,g_1) (h_2,e')(e,g_1)^{-1} \in H \times \{e'\}$
and
$(e,g_1) (e,g_2) \in \{e\} \times G$.

The usefulness and ubiquity of semidirect products has its roots in 
the following theorem. 
\begin{theorem}\label{ubi}
If $S$ is a group with a normal subgroup $H$ and a subgroup $G$ such
that 
\begin{enumerate}
\item
$H \cap G =\{ e \}$, where $e$ is the unit element of $S$, and 
\item 
every element $s \in S$ can be written as $s = hg$ with $h \in H$ and 
$g \in G$, 
\end{enumerate}
then the decomposition $s = hg$ is unique, $S/H \cong G$ and,
via $s=hg \to (h,g)$, the group $S$ is
isomorphic to $H \rtimes_\phi G$ with $\phi(g)h = ghg^{-1}$.
\end{theorem}
\noindent
\textbf{Proof:}
First we show the uniqueness of the decomposition of $s \in S$. Let
us assume that $s = hg = h'g'$ with $h,\, h' \in H$ and 
$g,\, g' \in G$. This assumption leads to the relation
${h'}^{-1} h = g' g^{-1}$ with the element on the left-hand side being
in $H$ and the element on the right-hand side being in $G$. Since the
intersection of $H$ and $G$ consists of the unit element only, we
find $h = h'$ and $g = g'$, \textit{i.e.} uniqueness of the
decomposition. The isomorphism between $S/H$ and $G$ is given by 
$Hg \leftrightarrow g$.
Now we assume that we have elements $s_1 = h_1g_1$ and 
$s_2 = h_2g_2$ of $S$. Then the isomorphism between $S$ and 
$H \rtimes G$ follows readily from 
\begin{equation}\label{mult}
s_1 s_2 = h_1 g_1 h_2 g_2 = h_1 (g_1 h_2 g_1^{-1}) g_1 g_2 \to 
(h_1 (g_1 h_2 g_1^{-1}), g_1 g_2) = (h_1,g_1) (h_2,g_2).
\end{equation}
Q.E.D.

We stress that in equation~(\ref{mult}) the mapping 
$\phi(g)h = ghg^{-1}$ of $G$ into $\mathrm{Aut}(H)$
comes about simply by the group
multiplication in $S$ and reordering of the factors in the product $s_1s_2$
in order to recover the decomposition of theorem~\ref{ubi}. This is 
completely analogous to the procedure leading to equation~(\ref{mult1}). 

\section{$\Delta(27)$, $\Delta(54)$ and tensor products of their
  three-dimensional irreps} 
\setcounter{equation}{0}
\renewcommand{\theequation}{B.\arabic{equation}}
\label{delta}

\paragraph{Introductory remarks:}
Before we discuss tensor products of $\Delta(27)$ and $\Delta(54)$,
some introductory remarks 
concerning $SU(3)$ subgroups are at order. Such a subgroup $G$ can also be
conceived as a three-dimensional representation which we denote generically by 
$\underline{3}$. Moreover, with $\underline{3}$ we automatically also have the
complex conjugate representation $\underline{3}^*$. 
Therefore, the $SU(3)$ relations for the tensor products
\begin{equation}\label{su3:3x3}
\underline{3} \otimes \underline{3}^* = \underline{1} \oplus \underline{8},
\quad
\underline{3} \otimes \underline{3} = \underline{3}^* \oplus \underline{6}
\end{equation}
hold. If we denote the Cartesian unit vectors of $\mathbbm{C}^3$ by $e_j$
($j=1,2,3$), the $\underline{1}$ is the trivial irrep acting on the basis
vector 
\begin{equation}
\frac{1}{\sqrt{3}} \left( e_1 \otimes e_1 + e_2 \otimes e_2 + e_3 \otimes e_3
\right),  
\end{equation}
while the $\underline{3}^*$ 
resides in the subspace spanned by the vectors
\begin{equation}\label{a}
a_j = \frac{1}{\sqrt{2}}\,\epsilon_{jkl} e_k \otimes e_l,
\end{equation}
where the $\epsilon_{jkl}$ are the components of the totally antisymmetric
$\epsilon$-tensor. 
Let us assume that $\underline{3}$ is irreducible. Then we make the following
useful observations:
\begin{itemize}
\item
For $G \subset SU(3)$ the $\underline{6}$ and the $\underline{8}$ may be
irreducible or not.
\item
If $G$ contains the center of $SU(3)$, then 
the $\underline{8}$ represents the center trivially.
\end{itemize}
A thorough discussion of tensor products of three-dimensional irreps of finite
subgroups of $SU(3)$ is found in~\cite{ludl}.

\paragraph{The tensor product $\three \otimes \three^*$:}
It is convenient to exploit the fact that this tensor product  
decays into nine one-dimensional irreps under
$\Delta(27)$, \textit{i.e.} 
one can choose a common basis of eigenvectors of 
$C \otimes C^*$ and $E \otimes E$:
\begin{equation}\label{b}
\renewcommand{\arraystretch}{1.2}
\begin{array}{ccc}
b_{00} & = &
\frac{1}{\sqrt{3}} \left(
e_1 \otimes e_1 + e_2 \otimes e_2 + e_3 \otimes e_3
\right), \\
b_{01} & = &
\frac{1}{\sqrt{3}} \left(
e_1 \otimes e_1 + \omega e_2 \otimes e_2 + \omega^2 e_3 \otimes e_3
\right), \\
b_{02} & = &
\frac{1}{\sqrt{3}} \left(
e_1 \otimes e_1 + \omega^2 e_2 \otimes e_2 + \omega e_3 \otimes e_3
\right), \\
b_{10} & = &
\frac{1}{\sqrt{3}} \left(
e_2 \otimes e_1 + e_3 \otimes e_2 + e_1 \otimes e_3
\right), \\
b_{11} & = &
\frac{1}{\sqrt{3}} \left(
e_2 \otimes e_1 + \omega e_3 \otimes e_2 + \omega^2 e_1 \otimes e_3
\right), \\
b_{12} & = &
\frac{1}{\sqrt{3}} \left(
e_1 \otimes e_3 + \omega^2 e_2 \otimes e_1 + \omega e_3 \otimes e_2
\right), \\
b_{20} & = &
\frac{1}{\sqrt{3}} \left(
e_1 \otimes e_2 + e_2 \otimes e_3 + e_3 \otimes e_1
\right), \\
b_{21} & = &
\frac{1}{\sqrt{3}} \left(
e_1 \otimes e_2 + \omega e_2 \otimes e_3 + \omega^2 e_3 \otimes e_1
\right), \\
b_{22} & = &
\frac{1}{\sqrt{3}} \left(
e_3 \otimes e_1 + \omega^2 e_1 \otimes e_2 + \omega e_2 \otimes e_3
\right)
\end{array}
\end{equation}
The eigenvalues under the action of $C$ and $E$ are given by 
\begin{equation}\label{eigenCE}
\left( C \otimes C^* \right) b_{pq} = \omega^p b_{pq}
\quad \mbox{and} \quad
\left( E \otimes E \right) b_{pq} = \omega^q b_{pq},
\end{equation}
respectively. 
Consequently, with equation~(\ref{1pq}) we arrive at the result
\begin{equation}\label{d27:3x3*}
\Delta (27): \quad
\three \otimes \three^* = 
\bigoplus_{p,q=0}^2 \one^{(p,q)}, 
\end{equation}
which nicely illustrates the decay of the $\underline{8}$ of
equation~(\ref{su3:3x3}) into irreps, 
in this case into eight one-dimensional irreps.

Moving to $\Delta(54)$, we have to apply $V^2$ to the basis
vectors~(\ref{b}) which has the effect that, apart from $b_{00}$, these basis
vectors are grouped in two to form the two-dimensional irreps: 
\begin{equation}
\two:   \; \{ b_{01},\, b_{02} \}, \quad
\two':  \; \{ b_{10},\, b_{20} \}, \quad
\two'': \; \{ b_{11},\, b_{22} \}, \quad
\two''':\; \{ b_{12},\, b_{21} \}.
\end{equation}
The labels for the irreps refer to equations~(\ref{2-dim-irrep}) and
(\ref{2-dim-irrep'}). Thus, we have just derived
\begin{equation}\label{d54:3x3*}
\Delta (54): \quad
\three \otimes \three^* = \one \oplus 
\two \oplus \two' \oplus \two'' \oplus \two'''.
\end{equation}

\paragraph{The tensor product $\three \otimes \three$:}
We complete the basis vectors~(\ref{a}) to a basis of 
$\mathbbm{C}^3 \otimes \mathbbm{C}^3$ by adding the sets of basis vectors
\begin{equation}
\{ e_1 \otimes e_1,\, e_2 \otimes e_2,\, e_3 \otimes e_3 \}
\end{equation}
and
\begin{equation}
\left\{ 
\frac{1}{\sqrt{2}} \left( e_2 \otimes e_3 + e_3 \otimes e_2 \right),\,
\frac{1}{\sqrt{2}} \left( e_3 \otimes e_1 + e_1 \otimes e_3 \right),\,
\frac{1}{\sqrt{2}} \left( e_1 \otimes e_2 + e_2 \otimes e_1 \right)
\right\}.
\end{equation}
On both sets, the action of tensor products of the group generators is given by
\begin{equation}
C \otimes C \to C^*, \quad E \otimes E \to E, \quad 
V^2 \otimes V^2 \to -V^2.
\end{equation}
Therefore, we obtain
\begin{equation}\label{d27:3x3}
\Delta (27): \quad
\three \otimes \three = \three^* \oplus 
\three^* \oplus \three^*
\end{equation}
and
\begin{equation}\label{d54:3x3}
\Delta (54): \quad
\three \otimes \three = \three^* \oplus 
\left( \three' \right)^* \oplus \left( \three' \right)^*
\end{equation}

For the irreps and the character table of $\Delta(27)$ see
section~\ref{D27}, for those of $\Delta(54)$ see section~\ref{D54}.
Of course, equations~(\ref{d27:3x3*}), (\ref{d54:3x3*}), (\ref{d27:3x3}) and
(\ref{d54:3x3}) could have also been derived from the corresponding character
tables.

\section{Tensor products of of three-dimensional irreps of $\saa$}
\setcounter{equation}{0}
\renewcommand{\theequation}{C.\arabic{equation}}
\label{tensor3}

\paragraph{The four-dimensional irreps of $\saa$:}
We denote the defining irrep~(\ref{def}) by 
$\three^{(0)}$. 
We know already from section~\ref{factor groups and irreps} that the
four-dimensional irreps are irreps of $\Sigma(36)$ which has a trivial
center. This suggests, according to the introduction of
appendix~\ref{delta}, to consider 
$\three^{(0)} \otimes \left( \three^{(0)} \right)^*$ in search
of these irreps.
We again use the basis~(\ref{b}) and the fact that 
it is a basis of eigenvectors of $C$ and $E$.
It remains to compute the action of $V \otimes V^*$ on $b_{pq}$.
The result of the tedious computation 
can graphically be presented in the following way:
\begin{equation}
V \otimes V^*: \quad 
\begin{array}{ccc}
b_{01} & \to & b_{20} \\
\uparrow && \downarrow \\
b_{10} & \leftarrow & b_{02}
\end{array}
\quad
\begin{array}{ccc}
b_{11} & \to & b_{21} \\
\uparrow && \downarrow \\
b_{12} & \leftarrow & b_{22}
\end{array}
\end{equation}
It is easy to check that the matrices of generators in the
four-dimensional irreps of equation~(\ref{four}) and (\ref{four'})
are obtained by the following ordering of the basis
elements:
\begin{equation}
\begin{array}{cc}
\four:        & \{ b_{01}, b_{10}, b_{02}, b_{20} \}, \\
\four^\prime: & \{ b_{11}, b_{12}, b_{22}, b_{21} \}. 
\end{array}
\end{equation}
Thus the decomposition of 
$\three^{(0)} \otimes \left( \three^{(0)} \right)^*$ is given by 
\begin{equation}\label{3x3*}
\three^{(0)} \otimes \left( \three^{(0)} \right)^* = 
\one^{(0)} \oplus \four \oplus \four^\prime.
\end{equation}

Taking the definition of the irreps $\three^{(p)}$ from
equation~(\ref{3dimirreps}), 
the relation~(\ref{3x3*}) can readily be generalized to
\begin{equation}
\three^{(p)} \otimes (\three^{(p')})^* = 
\one^{(p-p')} \oplus \four \oplus \four^\prime,
\end{equation}
where $p-p'$ has to be taken modulo 4.
In this equation we have used that
$\one^{(p)} \otimes \four \cong \four$ and 
$\one^{(p)} \otimes \four^\prime \cong \four^\prime$ ,
which can be read off from the character table because
$\chi_{\alpha k} \cdot \chi_{\beta k} = \chi_{\beta k}$ for all
conjugacy classes $C_k$
whenever $\alpha$ denotes a one-dimensional and $\beta$
a four-dimensional irrep. 

\paragraph{The tensor product $\three^{(0)} \otimes \three^{(0)}$:}
According to equation~(\ref{su3:3x3}), this tensor product contains a 
$\left( \three^{(0)} \right)^*$ which 
resides in the subspace spanned by the basis vectors~(\ref{a}).
Since $\saa$ has no six-dimensional irrep, the symmetric
part of the tensor product must decay into invariant subspaces.
To verify this statement, we choose an orthogonal system of vectors
\begin{equation}
\renewcommand{\arraystretch}{1.2}
\begin{array}{ccc}
\bar f_1 &=& e_1 \otimes e_1 + 
\zeta (e_2 \otimes e_3 + e_3 \otimes e_2 ), \\
\bar f_2 &=& e_2 \otimes e_2 + 
\zeta (e_3 \otimes e_1 + e_1 \otimes e_3 ), \\
\bar f_3 &=& e_3 \otimes e_3 + 
\zeta (e_1 \otimes e_2 + e_2 \otimes e_1 ),
\end{array}
\end{equation}
where $\zeta$ is a parameter.
Note that the action of $C$ and $E$ on the $\bar f_j$ can be
characterized as 
\begin{equation}
C \otimes C: \; \bar f_j \to 
\omega^{2(j-1)} \bar f_j \;\, (j=1,2,3),
\quad
E \otimes E: \; \bar f_1 \to \bar f_3 \to \bar f_2 \to \bar f_1,
\end{equation}
which shows that, on the basis vectors $\bar f_j$,
$C$ and $E$ are represented by $C^*$ and $E$, respectively.
Next we consider the action of $V$ on the $\bar f_j$, in order to
investigate if for specific values of $\zeta$ 
the $\bar f_j$ form the basis of an invariant space:
\begin{equation}\label{Vinv}
\left( V \otimes V \right) \bar f_j = M_{kj} \bar f_k.
\end{equation}
Writing
\begin{equation}
\bar f_j = \Gamma^j_{kl} e_k \otimes e_l
\quad \mbox{with} \quad 
\Gamma^1 = \left( \begin{array}{ccc}
1 & 0 & 0 \\ 0 & 0 & \zeta \\ 0 & \zeta & 0
\end{array} \right) \quad \mbox{etc.},
\end{equation}
equation~(\ref{Vinv}) reads
\begin{equation}
V \Gamma^j V^T = M_{kj} \Gamma^k.
\end{equation}
Evaluating this equation and using $V^T = V$, we find 
\begin{subequations}
\begin{eqnarray}
V \Gamma^1 V &=& 
-\frac{1}{3} \left(
\begin{array}{ccc}
1+2\zeta & 1-\zeta & 1-\zeta \\
1-\zeta & 1+2\zeta & 1-\zeta \\
1-\zeta & 1-\zeta & 1+2\zeta 
\end{array} \right) = 
M_{k 1} \Gamma^k, \\
V \Gamma^2 V &=& 
-\frac{1}{3} \left(
\begin{array}{ccc}
1+2\zeta & (1-\zeta) \omega & (1-\zeta) \omega^2 \\
(1-\zeta) \omega & (1+2\zeta) \omega^2 & 1-\zeta \\
(1-\zeta) \omega^2 & 1-\zeta & (1+2\zeta) \omega 
\end{array} \right) = 
M_{k 2} \Gamma^k, \\
V \Gamma^3 V &=& 
-\frac{1}{3} \left(
\begin{array}{ccc}
1+2\zeta & (1-\zeta) \omega^2 & (1-\zeta) \omega \\
(1-\zeta) \omega^2 & (1+2\zeta) \omega & 1-\zeta \\
(1-\zeta) \omega & 1-\zeta & (1+2\zeta) \omega^2 
\end{array} \right) = 
M_{k 3} \Gamma^k. 
\end{eqnarray}
\end{subequations}
It is straightforward to check that these equations lead to the
consistent solution
\begin{equation}\label{M}
M = -\frac{1+2\zeta}{3}
\left( \begin{array}{ccc}
1 & 1 & 1 \\ 1 & \omega^2 & \omega \\ 1 & \omega & \omega^2
\end{array} \right), 
\end{equation}
provided 
$1+2\zeta = (1-\zeta)/\zeta$.
This leads to the quadratic equation
\begin{equation}
2 \zeta^2 + 2 \zeta - 1 = 0,
\end{equation}
with the solutions
\begin{equation}\label{zeta}
\zeta_\pm = \frac{ -1 \pm \sqrt{3}}{2}.
\end{equation}
Therefore, our result is
\begin{equation}\label{M1}
-\frac{1+2\zeta_\pm}{3} = \mp \frac{1}{\sqrt{3}} 
\quad \Rightarrow \quad 
M = \left( \mp i V \right)^*.
\end{equation}
Comparing with equation~(\ref{3dimirreps}), we have derived
\begin{equation}\label{s36x3:3x3}
\three^{(0)} \otimes \three^{(0)} = 
\left( \three^{(0)} \right)^* \oplus \left( \three^{(1)} \right)^* 
\oplus \left( \three^{(3)} \right)^*,
\end{equation}
where the upper signs in equations~(\ref{zeta}) and (\ref{M1}) refer to
the $\left( \three^{(3)} \right)^*$ and the lower signs to the 
$\left( \three^{(1)} \right)^*$.
The corresponding basis vectors normalized to one are given by 
\begin{equation}
f^{(\pm)}_j =  \frac{\tau_\pm}{\sqrt{12}}\, e_j \otimes e_j \pm
\frac{1}{\tau_\pm} \left( e_k \otimes e_l + e_l \otimes e_k \right)
\quad \mbox{with} \; j \neq k \neq l \neq j
\;  \mbox{and} \;
\tau_\pm = \sqrt{2(3 \pm \sqrt{3}\,)},
\end{equation}
for $\left( \three^{(3)} \right)^*$ and 
$\left( \three^{(1)} \right)^*$, respectively. 
It is easy to check that these vectors, together with
the $a_j$ of equation~(\ref{a}), form an orthonormal basis for 
$\three^{(0)} \otimes \three^{(0)}$.

Equation~(\ref{s36x3:3x3}) can be generalized to
\begin{equation}
\three^{(p)} \otimes \three^{(p')} = 
\left( \three^{(p_1)} \right)^* \oplus \left( \three^{(p_2)} \right)^* 
\oplus \left( \three^{(p_3)} \right)^*
\end{equation}
with
\begin{equation}
p_1 = (-p-p')\,\mbox{mod}\, 4, \quad
p_2 = (-p-p'+1)\,\mbox{mod}\, 4, \quad
p_3 = (-p-p'+3)\,\mbox{mod}\, 4.
\end{equation}

\section{The nine-dimensional irreps of $\scc$}
\setcounter{equation}{0}
\renewcommand{\theequation}{C.\arabic{equation}}
\label{nine-216x3}

For the notation concerning the irreps of $\scc$
consult section~\ref{irreps scc}.
\begin{theorem}
The tensor product $\three^{(0)} \otimes \three^{(a)}$ establishes a
nine-dimensional irrep of $\scc$.
\end{theorem}
\noindent
\textbf{Proof:}
In order to proof this theorem we show that, for any non-zero vector 
\begin{equation}
x = \sum_{i,j=1}^3 c_{ij} e_i \otimes e_j \in 
\mathbbm{C}^3 \otimes \mathbbm{C}^3,
\end{equation}
by application of the representation operators of 
$\three^{(0)} \otimes \three^{(a)}$ to $x$ we obtain a 
set of vectors which spans the whole space.
The $e_i$ are the Cartesian basis vectors.
First we observe that the elements $C,\, E,\, V,\, D \in \scc$ 
are represented by 
\begin{equation}\label{CE3}
C \to C \otimes \bone, \quad E \to E \otimes \bone, \quad
V \to V \otimes R_a, \quad D \to D \otimes (R_aR_b),
\end{equation}
respectively, in $\three^{(0)} \otimes \three^{(a)}$.
The matrices $R_a$, $R_b$ are given in equation~(\ref{Rab}).
At least one of the $c_{ij}$ must be non-zero. Without loss of
generality we can assume that there is an index $k$ such that 
$c_{k1} \neq 0$. If this is not the case, 
by application of $D$ we can always achieve this.
In the next step we consider the vector 
\begin{equation}\label{EEE}
y \equiv 
\left[ (\bone + E + E^2) \otimes \bone \right] x = 
\sum_{j=1}^3 d_j u \otimes e_j
\quad \mbox{with} \quad 
d_j = \sum_{i=1}^3 c_{ij}, \quad 
u = \left( \begin{array}{c} 1 \\ 1 \\ 1 
\end{array} \right).
\end{equation}
Again, without loss of generality, we can assume 
$d_1 \neq 0$. The reason is that we can apply the operator $C$ or
$C^2$ to $x$ before we perform the operation~(\ref{EEE}). 
This would change $d_1 = \sum_{i=1}^3 c_{i1}$ to 
\begin{equation}
d_1 = \sum_{i=1}^3 \omega^{(i-1)} c_{i1} 
\quad \mbox{or} \quad
d_1 = \sum_{i=1}^3 \omega^{2(i-1)} c_{i1} .
\end{equation}
By assumption, not all three versions of $d_1$ can be zero at the same
time because otherwise $c_{11} = c_{21} = c_{31} = 0$ 
in contradiction to one $c_{k1}$ being non-zero.

Now we continue with the vector $y$ and $d_1 \neq 0$.
We note that $V u = -i\sqrt{3}\, e_1$. This allows to construct the vector
\begin{equation}
z_1 = \frac{i}{\sqrt{3}} (V \otimes R_a) y = 
d_1 e_1 \otimes e_1 - d_2 e_1 \otimes e_2 - d_3 e_1 \otimes e_3.
\end{equation}
Furthermore, application of 
\begin{equation}
\frac{1}{3} \left( \bone + C + C^2 \right) \otimes \bone 
\end{equation}
to $y$ generates the vector
\begin{equation}
z_2 = \sum_{j=1}^3 d_j e_1 \otimes e_j.
\end{equation}
Therefore, we have shown that, starting with $x$, the vector 
$e_1 \otimes e_1 = (z_1 + z_2)/(2d_1)$ is necessarily in the
representation space of $\three^{(0)} \otimes \three^{(a)}$. 
Then, repeated application of $D$ and $E$ to $e_1 \otimes e_1$ generates
a basis of $\mathbbm{C}^3 \otimes \mathbbm{C}^3$. Q.E.D.

One can ask the question how many inequivalent nine-dimensional irreps exist.
Because of 
\begin{equation}
\three^{(p)} \otimes \three^{(a)} = 
\left( \one^{(p)} \otimes \three^{(0)} \right) \otimes \three^{(a)} 
\cong 
\three^{(0)} \otimes \left( \one^{(p)} \otimes \three^{(a)}  \right)
\cong 
\three^{(0)} \otimes \three^{(a)}
\end{equation}
there are only two, namely 
$\nine \equiv \three^{(0)} \otimes \three^{(a)}$
and 
$\nine^* \equiv \left(\three^{(0)}\right)^* \otimes \three^{(a)}$.

\end{document}